\documentclass{aa}  

\usepackage{graphicx}

\usepackage{txfonts}

\usepackage[colorlinks=true,linkcolor=blue,filecolor=blue,urlcolor=blue,citecolor=blue]{hyperref}

\AtBeginDocument{\mathcode`v=\varv}
\usepackage[normalem]{ulem}

\begin{document}

        \title{Full non-LTE multi-level radiative transfer}
        
        \subtitle{I. An atom with three bound infinitely sharp levels}
        
        \author{T. Lagache
                \inst{1,}
                \inst{2}     
                \fnmsep\thanks{\email{tristan.lagache@univ-tlse3.fr}}  
                \and
                F. Paletou
                \inst{1,}
                \inst{2}         
                \and
                M. Sampoorna
                \inst{3}
        }
        
        \institute{Université de Toulouse, Observatoire Midi-Pyrénées,
          Cnrs, Cnes, Irap, Toulouse, France
          \and Cnrs, Institut de
          Recherche en Astrophysique et Planétologie, 14 av. E. Belin,
          31400 Toulouse, France
          \and Indian Institute of
          Astrophysics, Koramangala, Bengaluru 560034, India
        }
       \date{Received April 01, 2025; accepted May 20, 2025}
        
        \abstract
        {The standard nonlocal thermodynamic equilibrium (non-LTE)
          multi-level radiative transfer problem only takes into
          account the deviation of the radiation field and atomic
          populations from their equilibrium distribution.}
        {We aim to show how to solve for the full non-LTE (FNLTE)
          multi-level radiative transfer problem, also accounting for
          deviation of the velocity distribution of the massive
          particles from Maxwellian. We considered, as a first step, a
          three-level atom with zero natural broadening.}
        {In this work, we present a new numerical scheme. Its
          initialisation relies on the classic, multi-level approximate $\Lambda$-iteration (MALI) method for the standard non-LTE
          problem. The radiative transfer equations, the kinetic equilibrium equations for atomic
          populations, and the Boltzmann equations for the velocity distribution functions were
          simultaneously iterated in order to obtain self-consistent
          particle distributions. During the process, the observer's frame
          absorption and emission profiles were re-computed at every
          iterative step by convolving the atomic frame quantities
          with the relevant velocity distribution function.}
          {We validate our numerical strategy by comparing our
          results with the standard non-LTE solutions in the limit of
          a two-level atom with Hummer's partial redistribution in
          frequency, and with a three-level atom with complete
          redistribution. In this work, we considered the so-called cross-redistribution problem. We then show new FNLTE results for a
          simple three-level atom while evaluating the assumptions
          made for the emission and absorption profiles of the
          standard non-LTE problem with partial and cross-redistribution. }
        {}
        
        
        \maketitle
        
        \section{Introduction}
        \label{sec:intro}
        
        The standard multi-level non-LTE (non-local thermodynamic
        equilibrium) radiative transfer problem is at the very heart
        of the interpretation of spectra from astrophysical objects
        such as stars, circumstellar discs, molecular clouds,
        and so on \citep[see][]{HMbook}. While the standard approach,
        which has been continuously developed since the late 1960s, accounts for deviations of the
        radiation field and the atomic populations from their
        equilibrium distributions, it assumes velocity distribution of
        radiating atoms to be Maxwellian.

        Hereafter, we consider the problem of full non-LTE (FNLTE) radiative transfer, which mainly consists of the
        self-consistent calculation of the velocity distribution functions (VDFs) of
        massive particles with the radiation field; i.e. the
        distribution of photons in the atmosphere.  In that context,
        the main numerical burden is to work directly with atomic
        velocities and therefore solve the kinetic equilibrium
        equations for each velocity.
        
        The kinetic theory of particles and photons \citep[see][]{OxeniusBook} on which
        our approach is based also introduces an additional physical
        process of atomic velocity-changing collisions, which appear naturally
        in the Boltzmann equations. \citet[][see also; e.g. \citealt{HOSI,HOSII}]{Hubeny_Cooper1986} studied the consequences of these
        collisions
        on the radiative transfer problem in detail \citep[see
        also the discussion in Sect.~4 of][]{SPV24}.
The multi-level problem was originally formulated by
        \citet[][hereafter HOSI and
          HOSII, respectively; see also \citealt{Oxenius65,OxeniusBook}]{HOSI,HOSII} using a
        semi-classical picture.
        
        Until now, only the
                two-level FNLTE problem has been solved by \citet[][hereafter
                PSP23; see also \citealt{UBU}]{PSP23} for infinitely sharp
                levels, and by \cite{SPV24} for naturally broadened excited
                levels. In the FNLTE multi-level radiative transfer problem, the
        Boltzmann equation for the VDFs of excited atoms need to be
        solved simultaneously with the radiative transfer equation for
        all radiatively allowed transitions. Furthermore, at every
        iteration we also need to recompute all the macroscopic
        absorption and emission profiles by convolving the atomic
        profiles with the relevant VDFs. Clearly, this is a
        numerically very challenging problem. As a first step, we considered a simple atomic model,
        namely a three-bound-level atom with zero natural broadening,
        using most data from \cite{Avrett}.
        
        The FNLTE formalism for multi-level atoms naturally accounts
for both resonance and Raman scattering
        contributions\footnote{In a scattering transition from level
        $a \to b \to c$, if $a=c$ we have resonance scattering; if $a
        \ne c$ and $E_a, \,E_c < E_b$  it is Raman
        scattering.}. To incorporate these scattering mechanisms
--        including the effects of partial redistribution (PRD) for
        resonance scattering and cross-redistribution (a.k.a. XRD) for
        Raman scattering -- in the standard (in the sense that VDF
        deviations from Maxwellian are only partially considered)
        non-LTE multi-level problem an approximate approach was
        adopted; namely, the absorption profile for all transitions was
        assumed to be the usual Voigt (or Doppler) profile, and a
        simplified form of the emission profile derived by
        HOSII was used \citep[see e.g.][]{hubeny1985ETLA,
          Uitenbroek_1989, FBF, MALI_XRD}. We remark that the
        simplifying assumptions behind this approximate approach were
        detailed in \cite{hubeny1985ETLA}, and their validity
        remains to be evaluated. However, none of these assumptions are
        used in our more complete FNLTE approach.
        
        This paper is organised as follows. In
        Sect.~\ref{sec:FNLTE}, we present the formalism of the FNLTE
        transfer problem and apply it to the case of a three-bound-level atom with zero natural broadening in
        Sect.~\ref{sec:FNLTE3}. Our new numerical scheme is described
        in Sect.~\ref{sec:numerical}. It is validated against various
        benchmark solutions in Sect.~\ref{sec:validation} (and
        Appendix~\ref{app:App3}). Supplementary material, useful for
        understanding our tests, is presented in the
        Appendix~\ref{app:App2}. New results are presented in
        Sect.~\ref{sec:Results}. Finally, conclusions and
        future developments are discussed in
        Sect.~\ref{sec:conclusion}.
        
        \section{The FNLTE multi-level problem}
        
        \label{sec:FNLTE}
        
        Solving a multi-level radiative transfer problem requires the
        self-consistent resolution of a set of equations, each of them
        describing the distribution of particles of different nature:
        photons and massive particles (atoms, ions, molecules, or
        electrons). Here, we used the kinetic approach developed by
        \cite{OxeniusBook,Oxenius65}, which was extended by HOSI and HOSII
        and recently dubbed FNLTE by \cite{PP21}. In this
        formalism, the equations of the transfer problem are all
        kinetic equations.
        
        \subsection{The kinetic equations of photons}
        
        The distribution of photons associated with the
        transition from level $i$ of energy $E_i$, to a level $j$ of
        energy $E_j$, is the specific intensity $I_{ij}$. For an
        unpolarised, time-independent problem in a plane-parallel 1D
        geometry, this distribution depends on the frequency
        $\nu_{ij}$ of the photon in the observer’s frame, its
        direction cosine $\mu_{ij}=\cos(\theta_{ij}),$ and the optical
        depth $\tau_{ij}(\nu_{ij})$. Its kinetic equation is the usual
        radiative transfer equation (RTE), which is written, with the
        convention $i<j$, as follows:
        \begin{equation}
                \mu_{ij} \frac{\partial I_{ij}
                  (\nu_{ij},\mu_{ij},\tau_{ij})}{\partial \tau_{ij}} =
                I_{ij} (\nu_{ij},\mu_{ij},\tau_{ij}) -
                \frac{\eta_{ji}(\nu_{ij},\tau_{ij})}{\chi_{ij}(\nu_{ij},\tau_{ij})}\,.
                \label{eq:ETR}
        \end{equation}
        Hereafter, we will not display dependence of relevant quantities on the
                optical depth $\tau_{ij}$ without loss of clarity. In the RTE, we introduced the emissivity $\eta_{ji}$ and the
                absorption coefficient $\chi_{ij}$. They depend on the
        frequency, the optical depth, the direction of propagation
        and, implicitly, the VDFs of the massive particles,
        knowledge of which is indispensable for the computation of the emission
        and absorption profiles.

        In the following, we assume the isotropy of these
        quantities,\footnote{Taking into account their angular
        dependence would require us to consider angular
        redistribution.} and we write
        \begin{equation}
                \eta_{ji}(\nu_{ij})=\frac{h\nu_{0,ij}}{4\pi}n_jA_{ji}\psi_{ji}(\nu_{ij})\,,
                \label{eq:eta}
        \end{equation} 
        and
        \begin{equation}
                \chi_{ij}(\nu_{ij})=\frac{h\nu_{0,ij}}{4\pi}n_iB_{ij}\varphi_{ij}(\nu_{ij})\,,
                \label{eq:chi}
        \end{equation}
        where $A_{ji}$ and $B_{ij}$ are the Einstein coefficients
        associated, respectively, with spontaneous emission and
        radiative absorption; $\nu_{0,ij}$ is the frequency of the $i
        \leftrightarrow j$ transition, and $n_i$ and
        $n_j$ are the population densities of the $i$ and $j$
        excited levels. Strictly speaking, we should also consider the
        stimulated emission characterised by the Einstein coefficient
        $B_{ji}$. However, in most astrophysical cases
        \citep[with the exception of masers; see e.g.][]{Maser},
        $n_jB_{ji} \ll n_iB_{ij}$, and stimulated emission is
        therefore considered as `negative absorption'. Also, taking
        this additional phenomenon into account makes it much more
        difficult to obtain atomic frame absorption and emission profiles
        (see Sect.~5 of HOSI). Therefore, as a first step, we
        neglected stimulated emission completely. We also introduced the
        absorption profile $\varphi_{ij}$ and the emission profile
        $\psi_{ji}$ in the observer's frame defined as
        \begin{equation}
                \varphi_{ij}(\nu_{ij}) = \oint \frac{d\Omega_{ij}}{4\pi} \int f_i(\vec{v}) \alpha_{ij}(\xi_{ij})d^3\vec{v} \,,
                \label{eq:phi1}
        \end{equation}
        and
        \begin{equation}
                \psi_{ji}(\nu_{ij}) = \oint \frac{d\Omega_{ji}}{4\pi} \int f_j(\vec{v}) \beta_{ji}(\xi_{ji})d^3\vec{v} \,,
                \label{eq:psi1}
        \end{equation}
        where $\beta_{ji}$\footnote{The atomic emission profile was
        previously written, $\eta$ (see e.g. \citealt{OxeniusBook}
          or HOSI and HOSII), but we renamed it $\beta$
        in order to avoid any confusion with the emissivity.} and
        $\alpha_{ij}$ are the atomic emission and absorption profiles;
        $f_i(\vec{v})$ and $f_j(\vec{v})$ are the
        VDFs, normalised to unity
        for, respectively, atoms in the excited level $i$ and $j$;
        $\vec{v}$ is the atomic velocity; $\vec{\Omega}_{ij}$ is the
        direction of a photon of frequency $\nu_{ij}$ depending on azimuth $\phi_{ij}$
        and co-latitude $\theta_{ij}$; and $\xi_{ij}$ is the photon
        frequency in the atomic frame. The latter is directly linked to
        the frequency in the observer's frame by the Fizeau-Doppler
        relationships:
        \begin{equation}
          \xi_{ij} = \nu_{ij} - \frac{\nu_{0,ij}}{c} \vec{v} \cdot
          \vec{\Omega_{ij}} \, .
                \label{eq:Doppler}
        \end{equation}
        
        Solving the RTE and therefore the radiative transfer problem
        requires knowledge of the dynamic and radiative properties of
        the atoms: (i) their velocity distribution, $f_i$, (ii) the
        atomic densities, $n_i,$ and (iii) the atomic profiles,
        $\alpha_{ij}$ and $\beta_{ji}$. Each of these elements are
        discussed below.
        
        \subsection{The kinetic equations of massive particles}
        
        Massive particles constituting an atmosphere are also
        described, as photons, by a distribution (in velocity)
        governed by a kinetic equation. Apart from their
          nature, the only properties that distinguish
        each of these atoms are their speed and
        direction. Each of them is therefore
          characterised by its own VDF. The standard
        radiative transfer, however, deals with the self-consistent
        determination of the number density of the atoms -- i.e. the
        first moment of these VDFs -- together with the radiation
        field so that nothing is known about these VDFs.  On the
        contrary, our more complete FNLTE approach requires the
        self-consistent resolution of all the kinetic equations
        describing all the particles distributions.

        In the following, we assume that the temperature of the
        medium is sufficiently small so that $n_1 \gg n_2, n_3,
        ..$. In that case, atoms are so numerous on the fundamental
        level that their population varies very slightly with respect
        to the LTE value, and we consider that their
        distribution is Maxwellian. We only considered bound-bound
        radiative and collisional processes. For atoms
        in excited levels, let $F_i = n_i f_i$ be the distribution
        associated with the excited level $i \geq 2$. The kinetic
        equation describing their distribution is the Boltzmann
        equation \citep[we neglected the streaming of particles as
          argued by \citealt{Hubeny81}; for more details, please
          refer to][]{PP21}:
        \begin{equation}
                C[F_i] 
                =\left( \frac{\delta F_i}{\delta t} \right)_{\rm{rad.}} + \left(  \frac{\delta F_i}{\delta t}\right)_{\rm{inel.}} + \left(  \frac{\delta F_i}{\delta t}\right)_{\rm{v.c.c.}} = 0\,.
        \end{equation} 
        Here, $C$ is the collisional operator composed of
        three terms of different physical origins: a radiative term
        (collisions with photons), an inelastic term (collisions with
        free electrons), and a velocity-changing collision term,
        i.e. collisions between atoms \citep[this nomenclature was
          discussed in detail in Sect. 4 of][]{SPV24}.
        
        Let $N_i = n_i f_i d^3vdV$ be the number of atoms at level $i$
        contained in a volume $d^3vdV$ of the phase space. Per second,
        $N_iA_{ij}$ atoms spontaneously emit a photon and
        deexcite towards a level $j<i$, and $N_iB_{ij}J_{ij}(\vec{v})$
        atoms absorb a photon and are excited to a level $j>i$. In
        the phase space, these atoms are destroyed. Here, $J_{ij}(\vec{v})$
        denotes the partial scattering integral for atoms at level $i$ and velocity in the range of $(\vec{v},\vec{v}+d\vec{v}), $  which is defined as
        \begin{equation}
                J_{ij}(\vec{v}) = \int \alpha_{ij}(\xi_{ij})
                I_{ij} d\xi_{ij} \,.
                \label{eq:Jij1}
        \end{equation}
        Similarly, the number of atoms created at level $i$ from a
        level $j>i$ is $N_jA_{ji}$, and that from a level $j<i$ is $N_jB_{ji}J_{ij}(\vec{v})$. Finally, the total
        variation of the number density of atoms at level $i$ in a
        volume $d^3vdV$ of the phase space, i.e. the distribution $F_i
        = N_i / (d^3vdV)$, resulting from the radiative processes
        alone, is
        \begin{equation}
                \begin{split}
                        \left( \frac{\delta F_i}{\delta t}
                        \right)_{\mathrm{rad.}}  &= \sum_{j<i} \left(
                        n_j f_j B_{ji} J_{ij}(\vec{v}) - n_i f_i A_{ij} \right)
                        \\ &+ \sum_{j>i} \left( n_j f_j A_{ji} - n_i
                        f_i B_{ij} J_{ij}(\vec{v}) \right) \,.
                \end{split}
                \label{eq:rad_boltz}
        \end{equation}
        For inelastic collisions, we proceeded in an equivalent way by
        introducing the collisional excitation and deexcitation
        coefficients $C_{ij}$\footnote{The $C_{ij}$ depend
        on the electronic density, which should, in future works, also
        be determined self-consistently with the kinetic equations.}. We have:
        \begin{equation}
                \left( \frac{\delta F_i}{\delta t}
                \right)_{\mathrm{inel.}} = \sum_{j \neq i} (n_jf_j
                C_{ji} - n_if_iC_{ij})
                \label{eq:inel_boltz} \,.
        \end{equation}
        Velocity-changing collisions also destroy and create particles
        in phase space. The rate of these collisions is quantified by the parameter $Q_{V,i}$ for
          each atomic level (see \citealt
          {OxeniusBook}, HOSI, and \citealt{PP21}). We thus write:
        \begin{equation}
                \left(\frac{\delta F_i}{\delta
                  t}\right)_{\mathrm{v.c.c.}} = n_i Q_{V,i} \left[f^M
                  - f_i\right]
                \label{eq:vcc_boltz} \,,
        \end{equation}
        where $f^M$ is the Maxwell--Boltzmann velocity distribution
        defined as
        \begin{equation}
                f^M(\vec{v}) = \frac{1}{v_{\rm{th}}^3\pi^{3/2}} e^{-\vec{v\cdot
                  \vec{v}/v_{\rm{th}}^2}} \,,
        \end{equation}
        with $v_{\mathrm{th}}$ being the most probable velocity. The Boltzmann equation is finally written, for level $i,$ as
        \begin{equation}
                \begin{split}
                        & \sum_{j<i} \left( n_j f_j B_{ji} J_{ji}(\vec{v}) -
                  n_i f_i A_{ij} \right) + \sum_{j>i} \left( n_j f_j
                  A_{ji} - n_i f_i B_{ij} J_{ij}(\vec{v}) \right)\\ &+\sum_{j
                    \neq i} (n_jf_j C_{ji} - n_if_iC_{ij}) +n_i
                  Q_{V,i} \left[f^M - f_i\right] =0 \,.
                \end{split}
                \label{eq:Boltz} 
        \end{equation}
        
        As we can see from Eq.~(\ref{eq:Boltz}), there is a coupling
        between each of the VDFs $f_i$ in the Boltzmann
        equations. This is precisely what the standard non-LTE
        radiative transfer cannot deal with. We also note that the
        Boltzmann equations are formally kinetic equilibrium
        equations (KEE). Integrating over all velocities\footnote{By integrating over all the velocities, we will be missing all the
        	potential
        	effects that may originate from the velocity-changing collisional
        	process.}, we have the equilibrium between all the
        processes that populate and depopulate the atomic levels in a
        given space of volume $dV$. Hereafter, we refer to the
        following as integrated (and more usual) equations of the
          kinetic equilibrium or IKEE:
        \begin{equation}
                \begin{split}
                        & \sum_{j<i} \left( n_j B_{ji}
                  \mathcal{J}_{ji} - n_i A_{ij} \right) + \sum_{j>i}
                  \left( n_j A_{ji} - n_i B_{ij} \mathcal{J}_{ij}
                  \right)\\ &+\sum_{j \neq i} (n_j C_{ji} - n_i
                  C_{ij}) =0 \,,
                \end{split}
                \label{eq:KEE}
        \end{equation}
        where $\mathcal{J}_{ij}$ is the scattering integral, completely
        equivalent to the standard definition of $\bar{J}_{ij}$, and
        defined as
        \begin{equation}
                \mathcal{J}_{ij} = \int_{\vec{v}} f_i(\vec{v}) J_{ij}(\vec{v}) d^3\vec{v} \,. 
                \label{eq:J_curv}
        \end{equation}
        
        \subsection{Atomic emission and absorption profiles of a three-bound-level atom}
        \label{sec:prof_ato}
        
        In what follows, we used a model atom with three bound
        levels. Using phenomenological arguments, HOSI obtained the
        explicit form of the atomic profiles required to calculate the
        observer's frame profiles (see Eqs.~\ref{eq:phi1}
        and~\ref{eq:psi1}) and the partial scattering integral
        $J_{ij}(\vec{v})$ (see Eq.~\ref{eq:Jij1}). For the sake of clarity,
        we briefly recall their main arguments. For more details, we
        refer the reader to HOSI and HOSII.
                
        First of all, it is important to come back to the notion of
        natural population. For an ensemble of atoms of the
        same species, we are interested in a radiative transition from
        a level $b$ to a level $a$. Moreover, there is a given
        radiation field permeating the atmosphere, i.e. a given
        quantity of photons whose characteristics (frequencies and
        direction of propagation) are well known. We
        consider a particular physical process populating the level
        $b$. If the number of atoms created in atomic level $b$ is
        independent of the number of photons already present, then the
        $b \rightarrow a$ radiative transition will result in a new
        distribution of photons that is completely independent of the
        pre-existing photon distribution in the
        atmosphere. Thus, we can say that the physical process
        populating level $b$ leads to its natural population. This is
        the case for spontaneous emission, inelastic collisions, and
        velocity-changing collisions. However, radiative absorption
        depends on the radiation field and so does not naturally
        populate the atomic levels.
        
        An atomic profile characterises a specific radiative
        transition; say, $b \rightarrow a$. To find an expression for
        it, we need to study all the transitions and series of
        transitions that populate level $b$ and, ultimately, lead to
        the $b \rightarrow a$ radiative transition. Let $c
        \rightarrow b$ be a transition populating level $b$:
        
        \begin{enumerate}
                
                \item[(i)] If $c$ naturally populates level $b$,
                  then the resulting distribution of $b \rightarrow a$
                  will be independent of the resulting distribution of
                  $c \rightarrow b$. Hence, we need to consider the
                  physical phenomenon involving a single photon,
                  namely the $b^* \rightarrow a$ radiative transition
                  (where the notation $l^*$ means that level $l$ is
                  naturally populated). To describe it, we use the
                  generalised redistribution function
                  $r_{ba}(\xi_{ba})$ related to the probability that a
                  photon of frequency between $\xi_{ba}$ and
                  $\xi_{ba}+d\xi_{ba}$ is involved in the $b
                  \rightarrow a$ transition.
                
                \item[(ii)] If $c$ does not naturally populate level $b$
                  but is itself naturally populated, we
                  need to consider a two-photon phenomenon ($c^*
                  \rightarrow b \rightarrow a$). To handle it, we
                   define the conditional probability $j_{cba}(\xi_{ba})$;                  i.e. the probability that a photon of frequency
                  between $\xi_{ba}$ and $\xi_{ba}+d\xi_{ba}$ is
                  involved in the $b \rightarrow a$ transition,
                  knowing that a photon of frequency between
                  $\xi_{cb}$ and $\xi_{cb}+d\xi_{cb}$ was previously
                  involved in the $c^* \rightarrow b$ transition. This
                  quantity is defined as (see Eq. 4.4 of HOSI):
                \begin{equation}
                        j_{cba}(\xi_{ba}) = \frac{ \int I_{cb} r_{cba}(\xi_{cb},\xi_{ba}) d\xi_{cb}}{\int I_{cb} r_{cb}(\xi_{cb}) d\xi_{cb}} \,.
                        \label{eq:j_cba}
                \end{equation}
                In this equation, $r_{cba}(\xi_{cb},\xi_{ba})$ is the
                generalised redistribution function describing the
                probability that two photons whose frequencies are, respectively, between $\xi_{cb}, \xi_{cb}+d\xi_{cb}$
                and $\xi_{ba},\xi_{ba}+d\xi_{ba}$, are engaged in the
                $c^* \rightarrow b \rightarrow a$ transition (see Sect.~3
                of HOSI). To understand the meaning of
                $j_{cba}(\xi_{ba})$, we need to interpret each
                quantity probabilistically. The specific intensity
                $I_{cb}$ can be related to the so-called photon
                number density \citep[see e.g.][]{HMbook}. Thus, we can see the denominator
                term of Eq.~(\ref{eq:j_cba}) as the total number of
                photons involved in the $c \rightarrow b$ transition
                and the numerator term as the fraction of photons
                created by the $b \rightarrow a$ transition that
                actually comes from the $c \rightarrow b$ transition.
                
                \item[(iii)] If $c$ does not naturally populate the
                  $b$ level and is non-naturally populated by a third
                  transition $d^* \rightarrow c$, we need to
                  consider a phenomenon involving three photons in the
                  series of transitions $d^* \rightarrow c \rightarrow
                  b \rightarrow a$. To describe such a phenomenon, we
                  should use the conditional probability $j_{dcba}(\xi_{ba})$
                  defined analogously to Eq.~(\ref{eq:j_cba}).
                
                \item[(iv)] It is clear that the $d$ level itself can
                  be populated non-naturally by a given $e \rightarrow
                  d$ transition, and similarly for $e$ with an $f
                  \rightarrow e$ transition, and so on. Finally, when a photon is
                  involved in the $b \rightarrow a$ transition, there
                  will be a given probability
                  that it originally comes from one of the series of
                  transitions previously described. The atomic profile
                  is the probability density describing all these
                  possibilities simultaneously. It is defined as the
                  average of all the series of possible transitions
                  weighted by their occurrence probabilities.
                
        \end{enumerate} 
        
        For atoms with three levels, we therefore need to determine
        three absorption profiles -- $\alpha_{12},\, \alpha_{13}$, and
        $\alpha_{23}$ -- as well as three emission profiles:
        $\beta_{21},\, \beta_{31}$, and $\beta_{32}$. The $\alpha_{12}$
        and $\alpha_{13}$ profiles are the simplest to
        determine. Since stimulated emission has been neglected, the
        fundamental level is naturally populated. Consequently, the probability of
        occurrence of such a phenomenon is certain; only case (i)
        needs to be taken into account. We therefore have
        \begin{equation}
                \alpha_{12} = r_{12}(\xi_{12}) \,
                \label{eq:alpha_12}
        \end{equation}
        and
        \begin{equation}
                \alpha_{13}=r_{13}(\xi_{13}) \,.
                \label{eq:alpha_13}
        \end{equation}
        For $\beta_{21}$ and $\alpha_{23}$, we need to take into
        account all the series of transitions leading to level $2$. It
        can be naturally populated, and we denote
        $\mathrm{prob}(\rightarrow 2^*)$ as the relevant probability
        following HOSI and HOSII notations. This is case (i). It can
        also be populated non-naturally from the fundamental level,
        and we express $\mathrm{prob}(1^* \Rightarrow 2)$ as its
        probability. This is case (ii). Combining these two phenomena,
        we have (see Eqs.~4.3 and~4.6 of HOSI)
        \begin{equation}
                \alpha_{23} = \mathrm{prob}(\rightarrow 2^*) r_{23} + \mathrm{prob}(1^* \Rightarrow 2) j_{123} \,
                \label{eq:alpha_23}
        \end{equation}
        and
        \begin{equation}
                \beta_{21} = \mathrm{prob}(\rightarrow 2^*) r_{21} + \mathrm{prob}(1^* \Rightarrow 2) j_{121} \,.
                \label{eq:beta_21}
        \end{equation}
        For $\beta_{32}$ and $\beta_{31}$, we are interested in the
        processes populating level 3. We had to take into account
        the processes which do not naturally populate level $3$ from
        level $2$ and therefore treat the cascade $1^* \rightarrow 2
        \rightarrow 3 \rightarrow 1$ or $1^* \rightarrow 2 \rightarrow
        3 \rightarrow 2$ (case (iii)). We then have (see Eqs.
        4.7 and 4.8 of HOSI)
        \begin{equation}
                \begin{split}
                        & \beta_{31} = \mathrm{prob}(\rightarrow 3^*) r_{31} + \mathrm{prob}(1^* \Rightarrow 3) j_{131} \\
                        & + \mathrm{prob}(\rightarrow 2^* \Rightarrow 3) j _{231} + \mathrm{prob}(1^* \Rightarrow 2 \Rightarrow 3)j_{1231} \,
                \end{split}
                \label{eq:beta_31}
        \end{equation}
        and
        \begin{equation}
                \begin{split}
                        & \beta_{32} = \mathrm{prob}(\rightarrow 3^*) r_{32} + \mathrm{prob}(1^* \Rightarrow 3) j_{132} \\
                        & + \mathrm{prob}(\rightarrow 2^* \Rightarrow 3) j _{232} + \mathrm{prob}(1^* \Rightarrow 2 \Rightarrow 3)j_{1232} \,.
                \end{split}
                \label{eq:beta_32}
        \end{equation}
        For more details on the expressions of `$\mathrm{prob}$' probabilities, we refer the reader to HOSI or
        \cite{OxeniusBook}. Knowledge of these quantities will not be
        explicitly necessary later on. In this work, we only studied
        the simplified case of an atom with three infinitely sharp
        levels. In Sect.~\ref{sec:FNLTE3}, we will see how this simplifies the expression
        of the general atomic profiles given above and, consequently,
        the whole problem.
        
        \section{A three-level atom with infinitely sharp levels}
        \label{sec:FNLTE3}
        
        With the exception of the fundamental level, atomic levels are
        generally broadened in energy. In the theoretical
        developments of HOS presented above, no assumptions on the
        atomic model were made. Considering the same physical
        processes, the assumption of infinitely sharp levels does not
        degrade the problem, although it does
        simplify it. However, we warn the reader that this
          hypothesis is part of a heuristic approach and that the
          realistic study of an astrophysical atmosphere does not
          leave us the choice of the atomic model, it imposes it. In fact, it is the physical conditions in which
        the atoms are immersed and the intrinsic nature of the atoms
        that dictate the width of the atomic levels (broadened or
        infinitely sharp). For example, if the conditions under investigation
        lead to collisional broadening, these physical
        conditions impose an atomic model. Similarly, if we assume an
        atom such that levels $4$ and $5$ are naturally broadened, but
levels $2$ and $3$ are not (e.g. metastable levels), this atomic model must be adopted for a realistic
        study. Hereafter, we restrict ourselves
        to an atmosphere composed of atoms with three bound and infinitely
        sharp levels, neglecting stimulated emission and considering that
        the VDF associated with the fundamental level is Maxwellian ($f_1\equiv f^M$).

        \subsection{Atomic profiles}
        
        To calculate the atomic profiles of a three-level atom, we
        distinguish five phenomena that ultimately lead to the
        radiative transition under investigation: (i) absorption or emission
        of a single photon in a given $b \rightarrow a$ transition
        ($r_{ab}$ terms); (ii) resonance scattering, i.e. absorption
        and re-emission of a photon in a single line ($j_{aba}$
        terms); (iii) Raman scattering, i.e. absorption and
        re-emission of a photon in two distinct lines ($j_{cba}$
        terms); (iv) two-photon absorption processes ($j_{123}$
        term) and (v) three-photon processes ($j_{dcba}$ terms).
Since our levels are assumed to be infinitely sharp, the
        probability that a photon involved in a given $b \rightarrow
        a$ transition has a frequency $\xi_{ba} \neq \nu_{0,ba}$ is
        zero. We then write
        \begin{equation}
                r_{ba}(\xi_{ba})  = \delta (\xi_{ba} - \nu_{0,ba}) \,,
        \end{equation}
        where $\delta$ is the Dirac distribution. Resonance scattering
        was studied by \cite{hummer62_redistrib}, which obtained
        analytical expressions for the redistribution functions
        $r_{aba}(\xi_{ab},\xi_{ba})$. Later, \cite{Hubeny_2photons}
        obtained general expressions for all two-photon redistribution
        functions $r_{cba}(\xi_{cb},\xi_{ba})$. For atoms without
        natural broadening, we have, for the Raman or resonance
        scattering processes \citep[see Eq. 10.135 of][]{HMbook},
        \begin{equation}
                r_{cba}(\xi_{cb},\xi_{ba}) = \delta(\xi_{cb} - \nu_{0,cb})\delta(\xi_{ba}-\nu_{0,ba}) = r_{cb}(\xi_{cb})r_{ba}(\xi_{ba}) \,. 
        \end{equation}
        The conditional probabilities $j_{cba}(\xi_{ba})$ are then
        written as follows:
        \begin{equation}
                j_{cba}(\xi_{ba}) = \frac{ \int I_{cb} r_{cb}(\xi_{cb}) r_{ba}(\xi_{ba}) d\xi_{cb}}{\int I_{cb} r_{cb}(\xi_{cb}) d\xi_{cb}}
                = r_{ba}(\xi_{ba}) \,.
                \label{eq:jcba}
        \end{equation}
        In this study, we did not consider the two-photon absorption
        process. As discussed in \citet[][p 322-323]{HMbook}, this
        process is negligible in most astrophysical atmospheres. We
also neglected all three-photon processes for which
        redistribution functions were obtained by
        \cite{Hubeny_3photons} from quantum mechanical
        calculations. For $\alpha_{23}$, for example, the term
        proportional to $j_{123}$ is assumed to be equal to zero,
        and therefore $\alpha_{23} = \mathrm{prob}( \rightarrow 2^*)
        r_{23}$. We know that atomic profiles and generalised
        redistribution functions are normalised to unity because they
        are probability densities. Hence, to guarantee this
        condition, we renormalised $\alpha_{23}$, which gives
        $\alpha_{23}=r_{23}$. We proceeded in the same way when
        three-photon processes were neglected.
        
        Finally, for a three-level atom with infinitely sharp levels,
        each of the six atomic profiles established in
        Eqs.~(\ref{eq:alpha_12}) to~(\ref{eq:beta_32}) can then be
        written simply, $\forall j>i$, after renormalisation, as
        \begin{equation}
                \alpha_{ij}(\xi_{ij}) = r_{ij}(\xi_{ij}) =
                \delta(\xi_{ij} - \nu_{0,ij}) \,
                \label{eq:prof_ato_abs}
        \end{equation}
        and
        \begin{equation}
                \beta_{ji}(\xi_{ji}) = r_{ji}(\xi_{ji}) =
                \delta(\xi_{ji} - \nu_{0,ij}) \,.
                \label{eq:prof_ato_em}
        \end{equation}
        
        \subsection{The multi-level atom problem}
        
        Our new problem therefore involves five coupled distributions:
        three specific intensities, associated with each of the
        radiatively allowed transitions, and two velocity
        distributions associated with the two excited levels. We must
        now jointly solve all the RTEs (see Eq.~\ref{eq:ETR}), together
        with the IKEE (see Eq.~\ref{eq:KEE}) and the Boltzmann
        equations (see Eq.~\ref{eq:Boltz}). The set of IKEEs is
          redundant. So, for atoms with three levels, we chose to
        replace one of the three equations, namely the one associated
        with the fundamental density level $n_1$, by the
        conservation equation $n_1 + n_2 + n_3 =
        n_\mathrm{tot}$. Thus, solving the IKEE is equivalent to
        solving the system:
        \begin{equation}
                \left( \begin{matrix}
                        1 & 1 & 1 \\
                        E_{21} & D_2 & E_{23} \\
                        E_{31} & E_{32} & D_3
                \end{matrix} \right)
                \left(\begin{matrix}
                        n_1 \\ n_2 \\n_3
                \end{matrix}\right)
                =       \left(\begin{matrix}
                        n_{\mathrm{tot}} \\ 0 \\ 0
                \end{matrix}\right) \,,
                \label{eq:KEE_sys}
        \end{equation}
        with
        \begin{equation}
                E_{ij} = A_{ji} + B_{ji}\mathcal{J}_{ji} + C_{ji} \,
        \end{equation}
        and
        \begin{equation}
                D_i = - \sum_{j \neq i} (A_{ij} + B_{ij}\mathcal{J}_{ij} + C_{ij}) \,.
        \end{equation}
        In the above equations, we (numerically) assume that
        $A_{ij}=0$ for $i<j$ and $B_{ij}=0$ for $i>j$.

        Solving Boltzmann equations means solving a set of three equations, one of which is trivial ($f_1=f^M$). We then have
        \begin{equation}
                \left( \begin{matrix}
                        1 & 0 & 0 \\
                        e_{21} & d_2 & e_{23} \\
                        e_{31} & e_{32} & d_3
                \end{matrix} \right)
                \left(\begin{matrix}
                        f_1(\vec{v}) \\ f_2(\vec{v}) \\f_3(\vec{v})
                \end{matrix}\right)
                =       \left(\begin{matrix}
                        1 \\ -n_2 Q_{V,2} \\ -n_3 Q_{V,3}
                \end{matrix}\right) f^M(\vec{v}) \,,
                \label{eq:Boltz_sys}
        \end{equation}
        with
        \begin{equation}
                e_{ij} = n_j  ( A_{ji} + B_{ji}J_{ji}(\vec{v}) + C_{ji} )\,
        \end{equation}
        and
        \begin{equation}
                d_i = - n_i \left( Q_{V,i} + \sum_{j \neq i} (A_{ij} + B_{ij}J_{ij}(\vec{v}) + C_{ij}) \right) \,.
                \label{eq:d_i}
        \end{equation}
        
        The choice of the atomic and scattering models does not
        directly affect the form of these equations. However, their
        ingredients do depend explicitly on them:
        the absorption and emission profiles in the observer's frame
        $\varphi_{ij}$ and $\psi_{ji}$ and the partial scattering
        integrals $J_{ij}(\vec{v}),$ given, respectively, by
        Eqs.~(\ref{eq:phi1}),~(\ref{eq:psi1}),
        and~(\ref{eq:Jij1}). For infinitely sharp levels, the
        emission and absorption profiles take the form (see
        Appendix~\ref{app:App1})
        \begin{equation}
                \psi_{ji}(x_{ji}) = \frac{1}{2 \Delta_{ji}} \int_{\lvert x_{ji} \rvert}^{\infty} f_j(u) u du \,
                \label{eq:psi2}
        \end{equation}
        and
        \begin{equation}
                \varphi_{ij}(x_{ij}) = \frac{1}{2 \Delta_{ij}} \int_{\lvert x_{ij} \rvert}^{\infty} f_i(u) u du \,,
                \label{eq:phi2}
        \end{equation}
        together with
        \begin{equation}
                J_{ij}(\vec{u}) = \oint \frac{d\Omega_{ij}}{4\pi} \int \delta(x_{ij} - \vec{u} \cdot \vec{\Omega_{ij}}) I_{ij} dx_{ij}\,.
                \label{eq:Jij2}
        \end{equation}
        
        In the equations above, we introduce the reduced
        frequency $x_{ij} = (\nu_{ij} - \nu_{0,ij})/\Delta_{ij}$; the
        normalised velocity $\vec{u} = \vec{v}/v_{\mathrm{th}}$ \footnote{We now write $f^M(\vec{u}) = \pi^{-3/2}e^{-\vec{u} \cdot \vec{u}}$. The factor $v_{\rm{th}}^3$ disappears, in the integrals, with the change of variable $v \rightarrow u$.}; and the Doppler width $\Delta_{ij} =
        (\nu_{0,ij}/c)v_{\mathrm{th}}$. We also note that $f_i(u)$ is
        the modulus of the VDF $f_i(\vec{u})$, i.e. its
          integration over all the directions $\vec{\Omega}_u$ of the
          atoms. Finally, it is easy to show that $\varphi_{12}$ and
        $\varphi_{13}$ are given by the mere Doppler profile,
        \begin{equation}
                \varphi_{ij}(x_{ij}) = \frac{1}{\Delta_{ij}} \times
                \frac{1}{\sqrt{\pi}} e^{-x_{ij}^2} \,,
        \end{equation}
        because $f_1 = f^M$. A priori, all the other profiles will be
        different, and only a self-consistent resolution of the RTE,
        IKEE, and Boltzmann equations will make it possible to obtain an expression
        that is physically consistent with the FNLTE approach to
        radiative transfer.
        
        \section{Numerical strategy}
        \label{sec:numerical}
        
        In this section, we present our numerical strategy for solving
        the FNLTE problem for three infinitely sharp bound-level
        atoms. We first initialised the iterative process to the
        multi-level CRD solution with the MALI method of \citet[][hereafter the MALI-CRD method; see also \citealt{MALICRD_HAL}]{MALI}. Then, we successively updated each of the radiative quantities using the following very simple strategy:
        
        \begin{itemize}
                \item[0:] Initialise with a MALI-CRD solution,
                  which gives the population densities $n_1, \, n_2,$
                  and $n_3$ and therefore the CRD source functions.
                  
                \item[1:] For each of the transitions, calculate
                  the emission and absorption coefficients, $\eta_{ji}$, $\chi_{ij}$, using
                  Eqs.~(\ref{eq:eta}) and~(\ref{eq:chi}).
                  
                \item[2:] Calculate the source function $S_{ij}$:
                \begin{equation}
                        S_{ij}(x_{ij}) =
                        \frac{\eta_{ji}(x_{ij})}{\chi_{ij}(x_{ij})} \, .
                \end{equation}
                
              \item[3:] Knowing the source function at all optical
                depths, compute photon distributions $I_{ij}$ for each
                of the transitions.
                
                \item[4:] Compute the partial scattering integrals,                  $J_{ij}(\vec{u}),$ given by Eq.~(\ref{eq:Jij2}), using the current VDF (a Maxwellian is adopted for the first iteration; see also
                  Sect. 4 of PSP23).
                  
                \item[5:] Compute scattering integrals,
                  $\mathcal{J}_{ij}$, using Eq.~(\ref{eq:J_curv}).
                  
                \item[6:] Update all populations, $n_i,$ by solving
                  the IKEE system in Eq.~(\ref{eq:KEE_sys}).
                  
                \item[7:] Update the velocity distributions, $f_i,$ by
                  solving the system in Eq.~(\ref{eq:Boltz_sys}) at
                  each optical depth and atomic velocity $\vec{u}$.
                  
                \item[8:] Update $\varphi_{ij}$ and $\psi_{ji}$
                  profiles for each of the radiatively allowed
                  transitions using Eqs.~(\ref{eq:phi2})
                  and~(\ref{eq:psi2}), and return to Step 1.
                  
        \end{itemize}
        
        The output of this iterative process is a self-consistent
        determination of all the photon and massive particle
        distributions.
        
        \section{Validation}
        \label{sec:validation}
        
                \begin{figure}[t]
                        \includegraphics[width=\columnwidth]{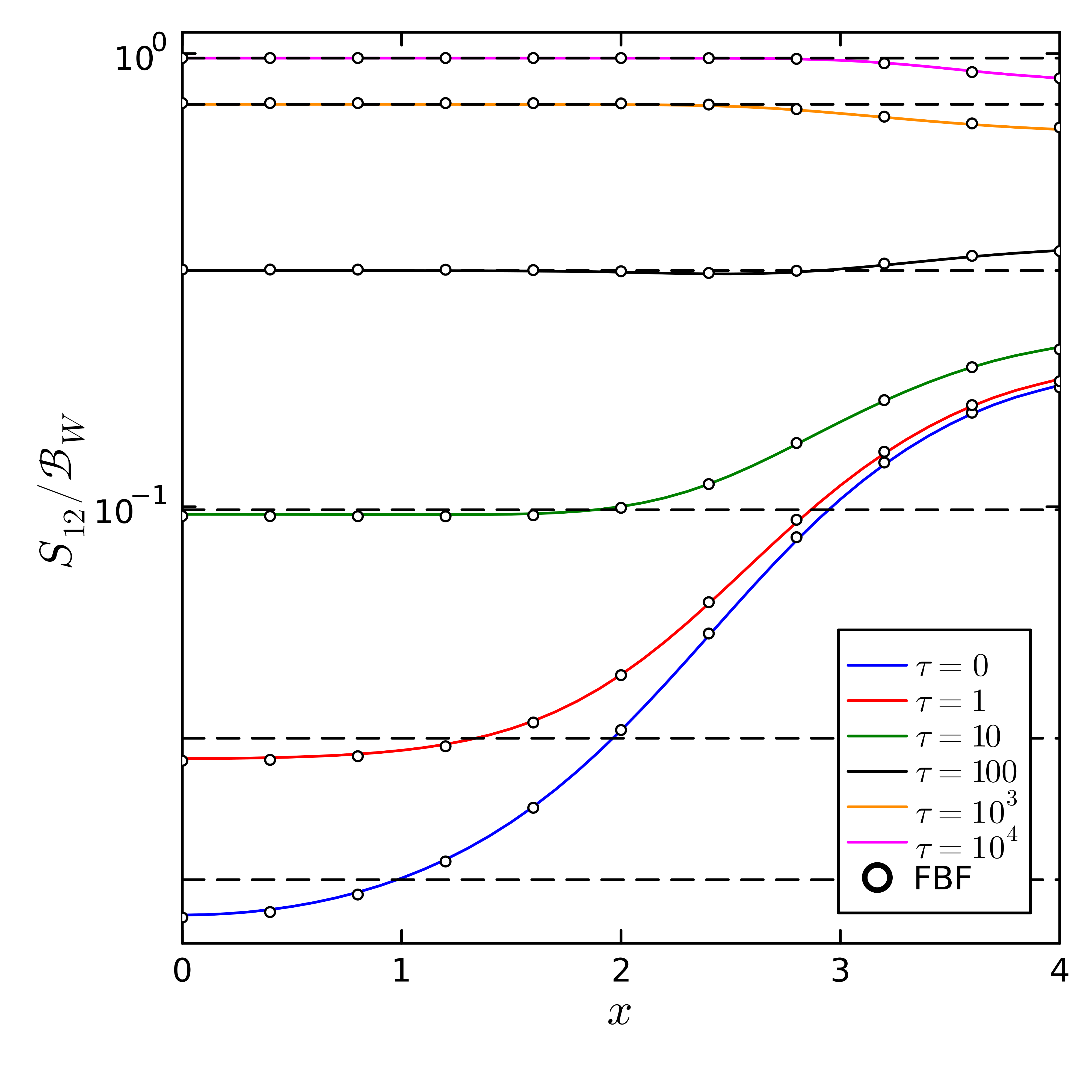}
                        \caption{Frequency variation of normalised source
                                function $S_{12}$ of $1\leftrightarrow 2$ line at different
                                optical depths $\tau$ for a two-level
                                atom. Horizontal dashed black lines show CRD
                                solutions for $\tau=0, 1, 10, 100, 10^3, 10^4$. We
                                compare FNLTE results (coloured lines) with Hummer's
                                PRD results (black open circle). CRD and PRD
                                solutions were computed using the ALI and FBF methods,
                                respectively. Clearly, we accurately recover the
                                results presented by \citet{hummer69_solutions} and
                                PSP23.}
                        \label{fig:Fig01_aa55008-25}
                \end{figure}
                
        Since the multi-level FNLTE radiative transfer problem was
        never fully addressed before, we proceeded with the
        validation of our calculations through several indispensable
        verification steps. Furthermore, since we considered the restricted case of a three-level atom with infinitely sharp levels and neglected velocity-changing collisions (see below), the numerical solutions presented here essentially validate our numerical scheme and do not represent the real spectral lines of hydrogen formed in a stellar atmosphere.
        
        All the solutions presented in this sections are calculated
        with a common set of parameters taken from \citet[][ Sect.~9.A; see also the code shared by \citealt{MALICRD_HAL}]{Avrett}\footnote{In particular, the temperature and collision rates $C_{ji}$ are fixed and independent of the optical depth.}. We
        considered an isothermal atmosphere composed of a three-bound-level
        hydrogen atom in a 1D, semi-infinite, plane-parallel geometry
        having a maximum total optical depth of $\tau_{\rm{max}} =
        10^{14}$ for the transition $1 \leftrightarrow 2$. We neglected the effects of
        velocity-changing collisions, i.e. $Q_V=0$ (however,
          $Q_V\neq 0$ adds no numerical difficulties).  We sample the optical depth scale with a logarithmic grid using four points per decade
        extending from $0$ to $\tau_{\mathrm{max}}$, with an initial
        step of $10^{-3}$. Integration over $\mu$ is performed using a
        six-point Gauss-Legendre quadrature. Integration over the
        azimuths (required to calculate $J_{ij}(\vec{u})$ -- see PSP23) was        performed using a ten-point rectangular
        quadrature. Finally, grids of reduced frequency, $x,$ and
        normalised velocity, $u,$ are identical and extend from $0$ to
        $x_{\rm{max}}=u_{\rm{max}}=4$ with a step of $0.1$.
        Velocity and frequency integrations were performed with
        trapezoidal quadrature. Each solution computed with our FNLTE
        code was obtained after $75$ iterations, and the process is
        always initialised by the CRD solution, obtained after $100$
        iterations of the MALI-CRD method. Finally, for quantitative
        comparison purposes, we introduce the following relative error
        term, defined for each optical depth and frequency,
        \begin{equation}
                \mathrm{Err}_{ij}^{\mathrm{rel}} = \left\lvert
                \frac{S_{ij}^{\mathrm{FNLTE}} -
                  S_{ij}^{\mathrm{ref}}}{S_{ij}^{\mathrm{ref}}}
                \right\lvert \,,
                \label{eq:Error term}
        \end{equation}
        between our FNLTE source functions $S_{ij}$ and reference solutions described hereafter.
                
                \begin{figure}[t]
                        \includegraphics[width=\columnwidth]{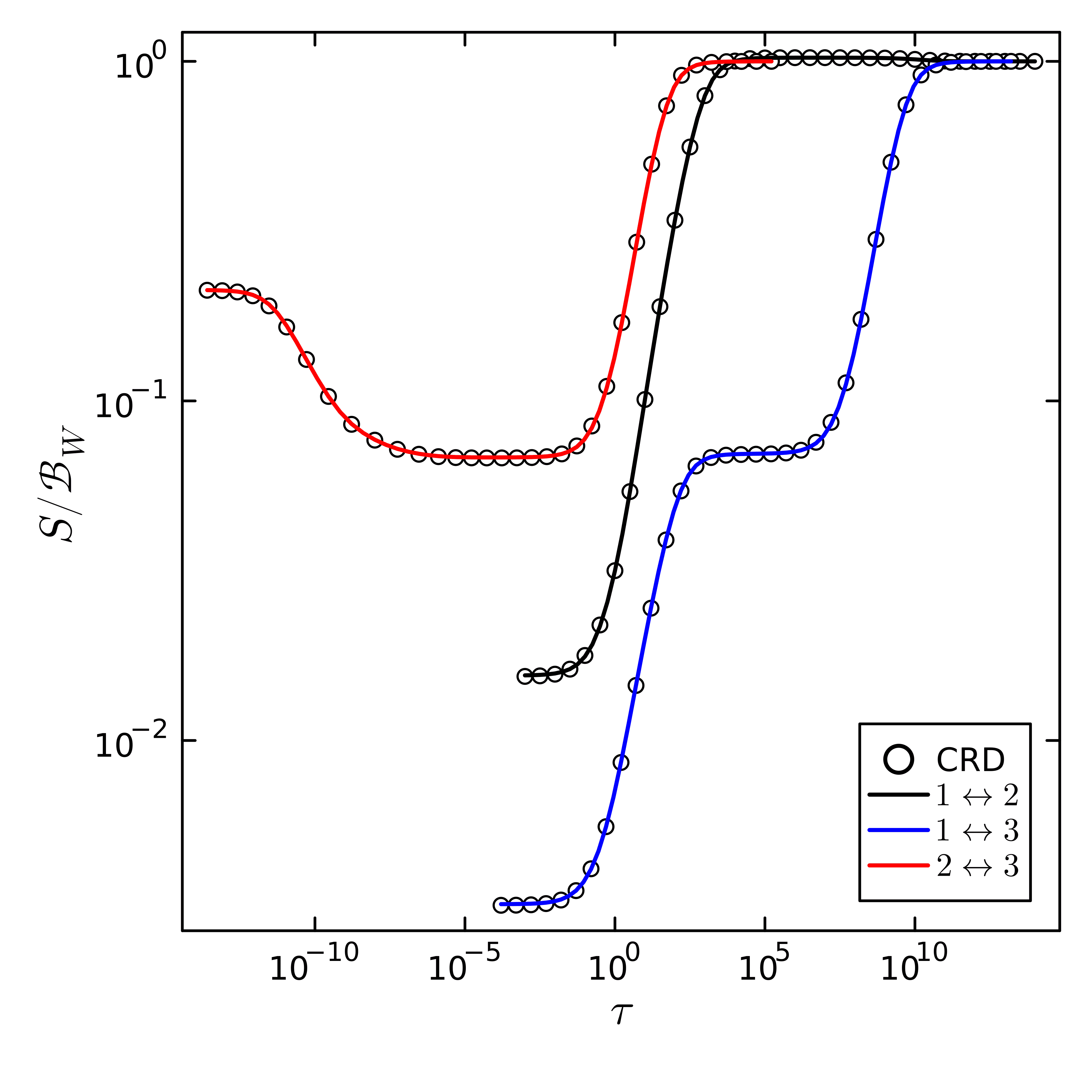}
                        \caption{Optical depth variations of normalised source
                                functions for each line of a three-level hydrogen atom
                                in CRD. Our three-level FNLTE code (coloured lines)
accurately reproduces reference solutions (open
                                black circle) computed with MALI-CRD.}
                        \label{fig:Fig02_aa55008-25}
                \end{figure}
                
        \subsection{Benchmarking against a two-level atom in PRD}
        
        The first step in validating our multi-level code is to reduce
        it to a two-level atom problem with coherent scattering in the
        atom's frame. PSP23 showed that when $Q_V=0$ it is equivalent to Hummer's results for PRD $R_{I-A}$
        \citep{hummer62_redistrib}. Fig.~\ref{fig:Fig01_aa55008-25} displays
        the source functions for $1 \leftrightarrow 2$ line, obtained with our FNLTE
        method (coloured lines) and compared to the reference $R_{I-A}$
        PRD solution (open black circles) calculated using the FBF method
        (\citealt{FBF}). The mean of the relative error between these
        solutions is $\approx 0.31\%$. We are also able to reproduce
        the results obtained by PSP23 for the $f_2$ distribution
        function.
        
        \subsection{Benchmarking against multi-level CRD}
        
        Next, by forcing, in our new three-level FNLTE code, that
        $f_1=f_2=f_3=f^M$, we recover the well-known results of
        MALI-CRD for Avrett's hydrogen case \citep[][see also \citealt{Paletou_Leger2007}]{Avrett}.
In Fig.~\ref{fig:Fig02_aa55008-25}, we show the optical depth
        variations of the source function calculated for each of the
        three lines using our multi-level FNLTE code and the MALI-CRD
        method. Mean relative errors of about $0.12 \%$ for $1 \leftrightarrow 2$,
        $0.33 \%$ for $1 \leftrightarrow 3,$ and $0.32 \%$ for $2 \leftrightarrow 3$ lines are
        reported, and they more likely come from completely different
        numerical schemes.
        
        \subsection{Benchmarking against XRD}
        \label{sec:approx_XRD}
        
                \begin{figure}[t]
                \includegraphics[width=\columnwidth]{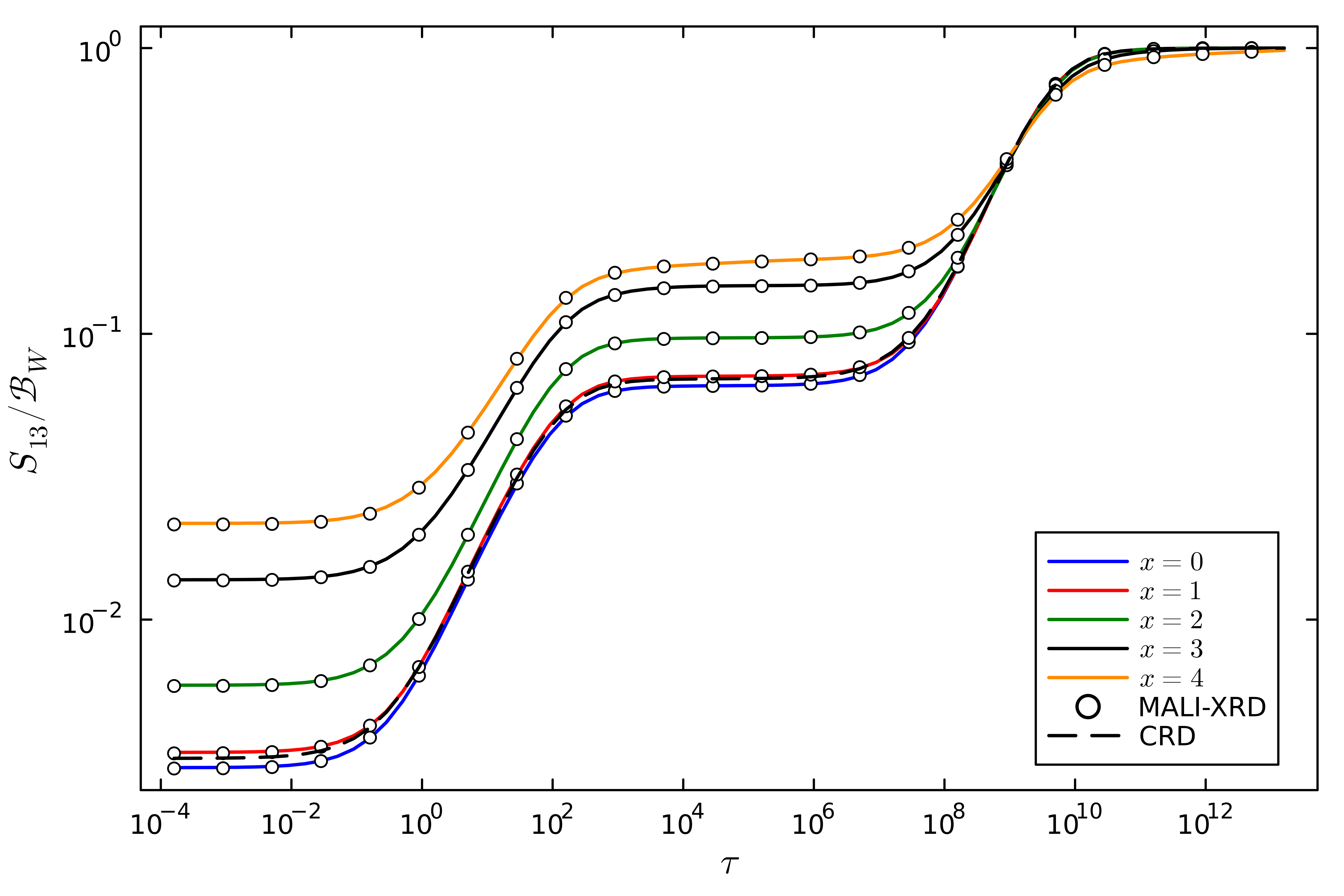}
                \caption{Optical depth variations of normalised source
                        functions for the $1 \leftrightarrow 3$ line of a three-level
                        hydrogen atom in the frame of XRD approximation. We
                        compare FNLTE results (coloured lines) for various
                        frequencies $x=0,1,2,3,4$ with XRD reference
                        solutions (open black circle) computed using the
                        MALI-XRD method of \cite{MALI_XRD}. The dashed
                        black line shows the CRD solution computed using the MALI-CRD
                        method.}
                \label{fig:Fig03_aa55008-25}
        \end{figure}
        
        On the basis of the theoretical work of HOSI \& HOSII, the
        FNLTE approach of radiative transfer presented in
        Sect.~\ref{sec:FNLTE} appears, at present, as the most
        complete approach (however, the QED approach of
          Bommier 2016 goes beyond the semi-classical picture of HOS
          formalism used here). In the past, indeed, numerous
          attempts have been made to incorporate it in more
          advanced radiative models, but with several simplifying
        assumptions. As a result of this, none of them could solve all
        the kinetic equations self-consistently with the RTE. The case
        where only resonance scattering is taken into account, namely
        the case of standard PRD, was studied by
        \cite{hubeny1985ETLA}. When Raman scattering is taken into
        account, we speak of `cross-redistribution' (this term
          was used by \citealt{milkey1975resonance} to describe this scattering
          process) or XRD \citep[e.g.][]{hubeny_lites,MALI_XRD,Uitenbroek_1989}. In all
        cases, these approaches were incomplete, with persisting
        approximations: (i) all absorption profiles and (ii) scattering integrals
        $\mathcal{J}_{ij}$ were calculated
        using Maxwellian VDFs, (iii) $d_i \approx n_i (D_i - Q_{V,i})$\footnote{Without v.c.c., we have $n_iD_i = \int d_i(\vec{u}) d^3\vec{u}$. Assuming $d_i\approx n_iD_i$ is equivalent to assuming that $\int d_i(\vec{u})f(\vec{u})d^3\vec{u} \approx \int d_i(\vec{u}) d^3\vec{u} \times \int f(\vec{u}) d^3\vec{u}$.} in
        the Boltzmann equations, and (iv)
        there is no coupling between the VDFs; in other words,
        to calculate $f_i$, we assumed that $f_j = f^M$ $\forall j \neq
        i$. Considering these assumptions, the Boltzmann
        equations presented in Eq.~(\ref{eq:Boltz_sys}) reduce to (assuming $Q_V=0$ again), for all atomic excited levels,
        \begin{equation}
                f_2 \approx \frac{f^M}{n_2 P_2} \Bigg\{
                \left[n_3(A_{32}+C_{32}) + n_1C_{12}\right] +
                n_1B_{12}J_{12}(\vec{u}) \Bigg\} \,
                \label{eq:Boltz_2_XRD}
        \end{equation}
        and
        \begin{equation}
                f_3 \approx \frac{f^M}{n_3P_3}\Bigg\{ \left[ n_1C_{13} +
                  n_2C_{23} \right] + n_1B_{13}J_{13}(\vec{u}) +
                n_2B_{23}J_{23}(\vec{u}) \Bigg\} \,,
                \label{eq:Boltz_3_XRD}
        \end{equation}
        where $P_i$ are defined as
        \begin{equation}
                P_i = \sum_{j \neq i} (A_{ij} + B_{ij}\mathcal{J}_{ij} + C_{ij}) \,.
        \end{equation}
        
        From these analytical expressions, we can deduce a general
        expression for the emission profiles (see
        Appendix~\ref{app:App2} for more details):
        \begin{equation}
                \begin{split}
                        \psi_{ji} = & \frac{\varphi_{ij}^{*M}}{n_j
                          P_j} \left\{ n_i B_{ij} \bar{J}_{iji}^M +
                        \sum_{\substack{l\neq i,j \\ l<j }} n_l B_{lj}
                        \bar{P}_{lji}^M \right. \\ & \left. + \left[
                          n_iC_{ij} +\sum_{k \neq i,j} n_k (A_{kj} +
                          C_{kj}) \right]
                        \vphantom{\sum_{\substack{l\neq i,j \\ l<j }}}
                        \right\} \,,
                \end{split}
                \label{eq:prof_em_general_formula_samp}
        \end{equation}
                \begin{figure}[t]
                \includegraphics[width=\columnwidth]{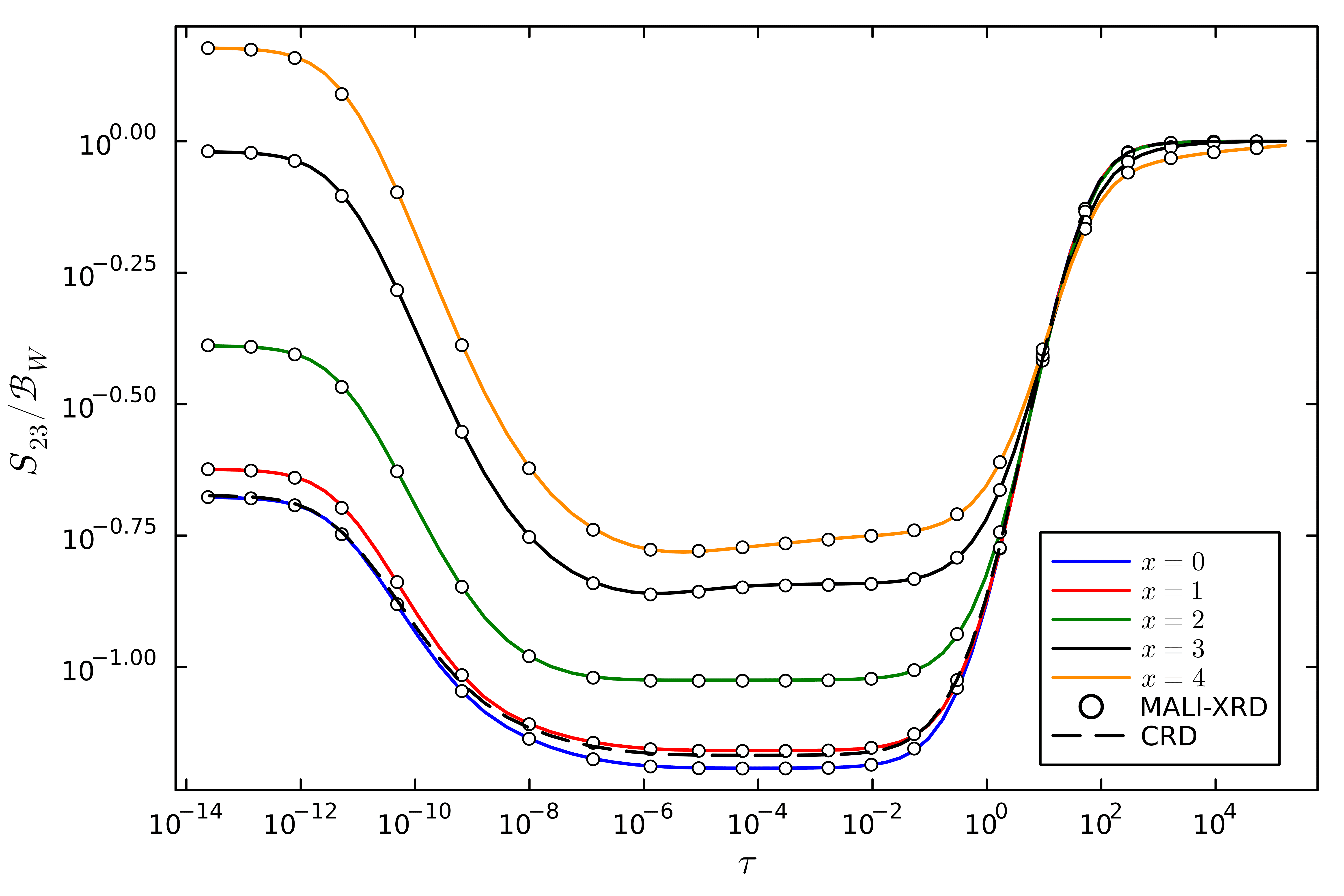}
                \caption{Same as Fig.~\ref{fig:Fig03_aa55008-25}, but for $2 \leftrightarrow 3$ line.}
                \label{fig:Fig04_aa55008-25}
        \end{figure}
        where the notation $X^M$ means that quantities were calculated
        using a Maxwellian VDF, in particular for the generalised
        redistribution functions, $R_{ijk}$, contained in the definition
        of the diffusion integrals $\bar{J}_{iji}$ and
        $\bar{P}_{lji}$. This last equation is completely equivalent
        to Eq.\,(11) of \cite{MALI_XRD} obtained in XRD;
        i.e. considering resonance and Raman scattering
        effects. Whether one works directly with the emission profiles
        (XRD) or solves the uncoupled kinetic equations (namely Eqs.~\ref{eq:Boltz_2_XRD}
        and~\ref{eq:Boltz_3_XRD}) self-consistently with the
        transfer equations, respective results should be identical. We
        show that, within the framework of the approximations
        presented above, there is an equivalence between our FNLTE
        formalism and the XRD formalism used by \cite{MALI_XRD}, whose
        numerical implementation is inspired by the MALI-PRD
        method developed by \cite{MALI_PRD}.
        
        The source functions obtained with these two methods are shown
        in Figs.~\ref{fig:Fig03_aa55008-25} and~\ref{fig:Fig04_aa55008-25}, respectively,
        for the $1 \leftrightarrow 3$ and $2 \leftrightarrow 3$ lines. For the $1 \leftrightarrow 3$ line,
        we measure a mean relative error between our
          degraded FNLTE code solution and the XRD solution of
        about $0.23\%$. For $2 \leftrightarrow 3$, it is about $0.37\%$. We
        chose not to show the source function associated with the
        $1 \leftrightarrow 2$ line, as it differs only slightly from the solution
        shown in Fig.~\ref{fig:Fig01_aa55008-25} (a relative maximum difference of $4.6
        \%$  with respect to the two-level atom
        solution). However, XRD results are very well reproduced
        for this line with our FNLTE code, with a mean relative
        error between these two solutions of about $0.23 \%$.
        
        In Fig.~\ref{fig:Fig04_aa55008-25}, at the surface and at high
        frequency, we observe an overshoot; i.e. $S_{23}/\mathcal{B}_W
        > 1$. The possibility that we are facing a numerical problem
        has been explored, in particular because we explored regions where
        the optical depths become very tiny. However, this feature is
        reproduced identically by the two different numerical
        methods. Furthermore, as the numerical problems remain
        identical, no similar behaviour has been observed in any of
        the other cases studied (such as MALI-CRD and MALI-PRD; the latter
        is discussed in Appendix~C). This characteristic seems
          to be specific to XRD, which suggests that the consistency
          of some of its assumptions would need to be discussed
          further. However, as we see in
        Sect.~\ref{sec:Results}, this artefact disappears when one finally deals
        with the complete problem.
The above-mentioned -- and very satisfactory -- tests now
          lead us with confidence towards new results, using the full description.
        
        \section{New results}
        \label{sec:Results}
        
        \begin{figure}
                \includegraphics[width=\columnwidth]{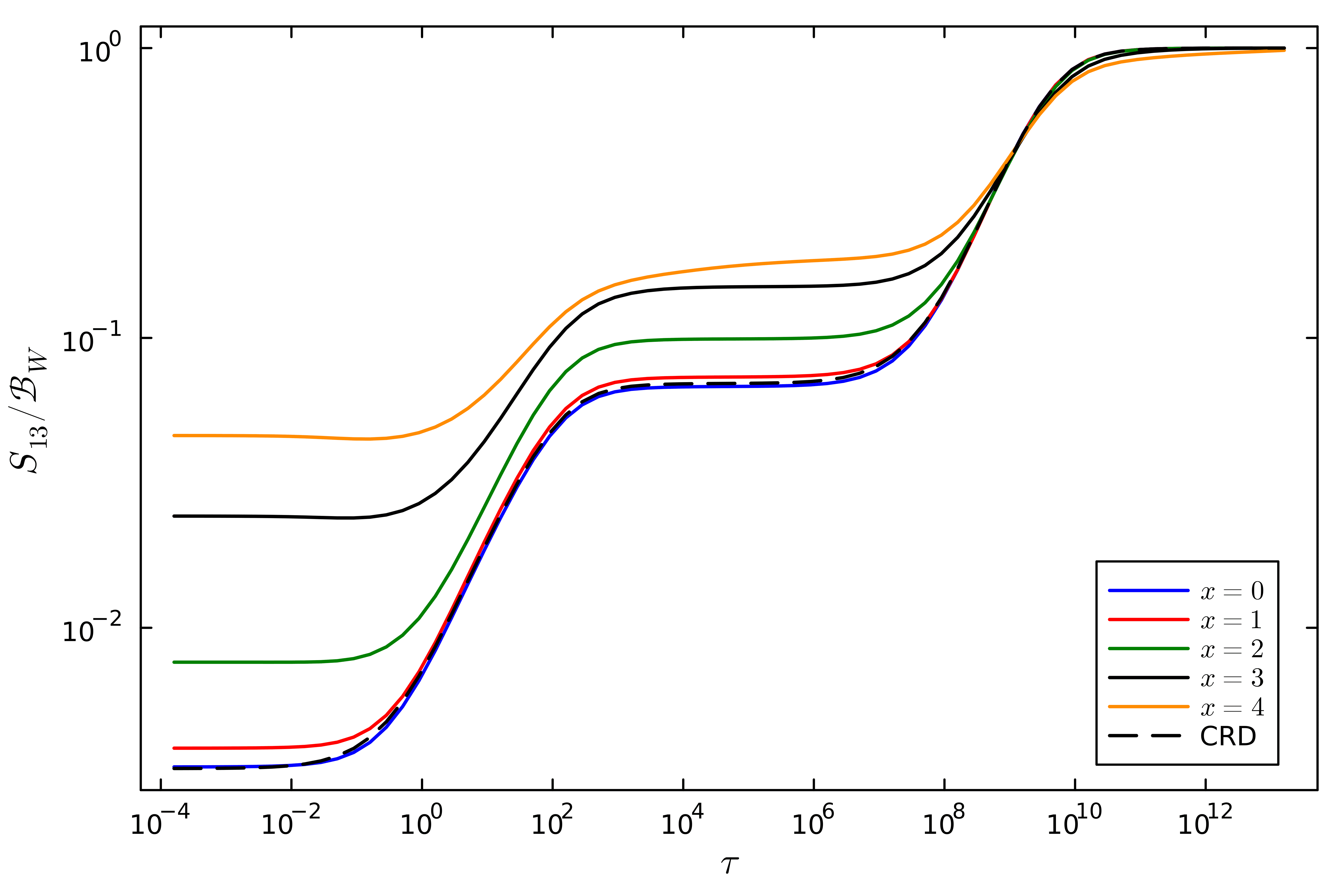}
                \caption{Optical depth variations of normalised source
                  functions for $1 \leftrightarrow 3$ line of a three-bound-level hydrogen atom. It shows FNLTE results
                  (coloured lines) obtained for various frequencies:
                  $x=0,1,2,3,4$. The dashed black line represents the
                  MALI-CRD solution.}
                \label{fig:Fig05_aa55008-25}
        \end{figure}
        
        \begin{figure}
                \includegraphics[width=\columnwidth]{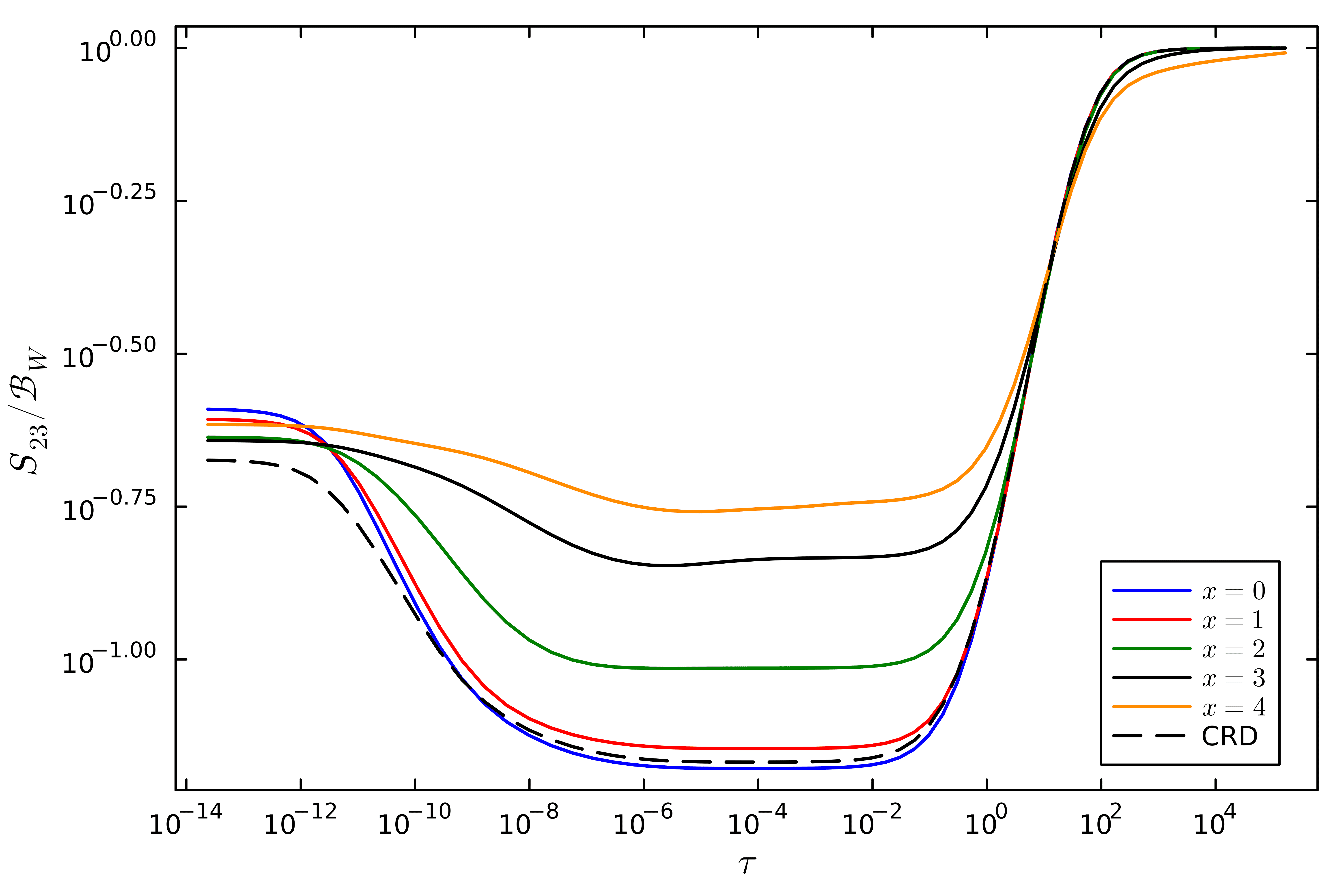}
                \caption{Same as Fig.~\ref{fig:Fig05_aa55008-25}, but for $2 \leftrightarrow 3$ line.}
                \label{fig:Fig06_aa55008-25}
        \end{figure}
        
        Hereafter, we used the same model and parameters as the ones
        used for our previous tests; in particular, velocity-changing
        collisions continue to be neglected (although their treatment
        does not present any numerical additional difficulty).
Firstly, the $S_{12}$ source function for the $1 \leftrightarrow 2$ line
        changes very slightly compared to the results obtained by
        PSP23 with a two-level atom with a mean relative error between
        these solutions of the order of $1.92\%$.  This can be
        explained by the fact that atomic levels are much less populated
        when their energy is higher. In other words, we have $n_1 \gg
        n_2 \gg n_3 \gg ..$. Thus, the source function for the $2
        \leftrightarrow 1$ transition is largely dominated by
        radiative processes populating level $2$ from level
        $1$ and vice versa. The impact of multi-level modelling on this
        line is therefore marginal.
                \begin{figure*}[t!]
                \centering
                \resizebox{375px}{232.5px}
                {\includegraphics[]{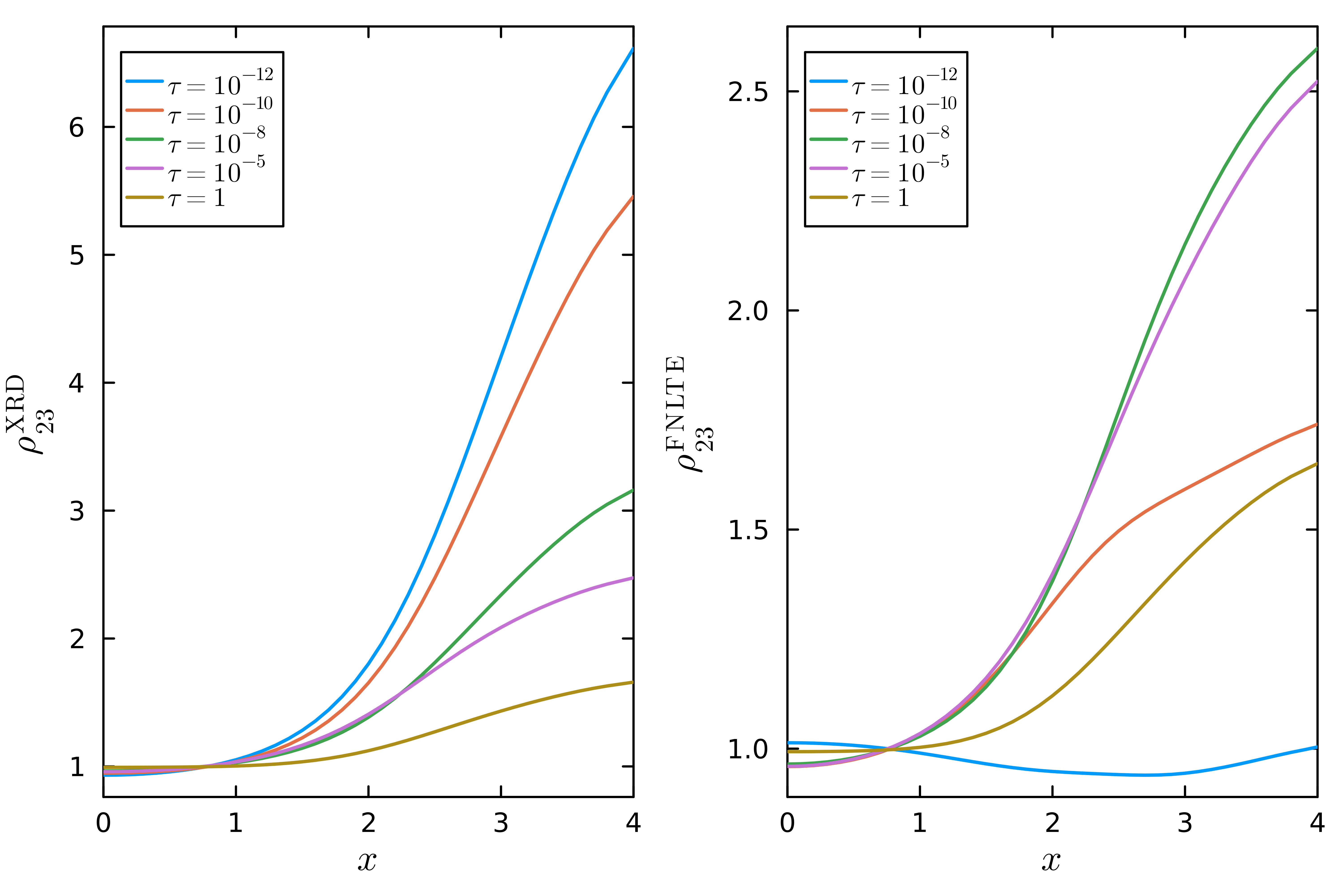}}
                \caption{Frequency variations of $\rho_{23}$ ratio
                        for $2 \leftrightarrow 3$ line of a three-bound-level hydrogen
                        atom. We show the XRD (left plot) and FNLTE (right
                        plot) results obtained for different optical depths:
                        $\tau=10^{-12},10^{-10},10^{-8},10^{-5},1$.}
                \label{fig:Fig07_aa55008-25}
        \end{figure*}
        On the other hand, the $1 \leftrightarrow 3$ and $2 \leftrightarrow 3$ lines are
        substantially affected by FNLTE effects. In
        Figs.~\ref{fig:Fig05_aa55008-25} and~\ref{fig:Fig06_aa55008-25}, we show the
        $S_{13}$ and $S_{23}$ source functions obtained for these
        lines. Differences can be seen at the surface with XRD,
        comparing Figs.~\ref{fig:Fig05_aa55008-25} and~\ref{fig:Fig06_aa55008-25},
        respectively, with Figs.~\ref{fig:Fig03_aa55008-25}
        and~\ref{fig:Fig04_aa55008-25}. We know that it is at the surface that
        non-LTE effects are more pronounced. It is therefore not
        surprising that any change in the approach, describing
        all deviations from LTE, modifies the characteristics of
        the atmosphere in this region. As we explain in the previous
        section, the overshoot observed using XRD for the $S_{23}$
        source function now disappears. This is because the frequency
        variations of the source function are mainly due to the ratio
        between the emission and the absorption profile:
        \begin{equation}
                \rho_{ij}(x) =
                \frac{\psi_{ji}(x)}{\varphi_{ij}(x)} \,.
        \end{equation}
        With XRD, we have $\rho_{23} \gg 1$ precisely in this region
        close to the surface and far from the line centre where the
        overshoot is observed, as can be seen from the left panel of
        Fig.~\ref{fig:Fig07_aa55008-25}. According to XRD assumptions, all
        absorption profiles are calculated using a
        Maxwellian. However, we see in Sects.~\ref{sec:FNLTE}
        and~\ref{sec:FNLTE3} that the absorption profile
        $\varphi_{23}$ should also be determined in a self-consistent
        way with all the other radiative quantities; i.e. calculated
        with $f_2 \neq f^M$. It therefore seems contradictory to use
        $f_2$ to determine the emission profile, $\psi_{21}$, but ignore
        this fact for the absorption profile, $\varphi_{23}$. This
        suggests that XRD's assumption about absorption profiles is
        largely responsible for the overshoot. This characteristic
        is no longer present when $\varphi_{23}$ is self-consistently
        calculated, and, quantitatively, we can see that $\rho_{23}$
        decreases significantly in the very same regions (see the
        right panel of Fig.~\ref{fig:Fig07_aa55008-25}).  
        
        A more detailed examination of the influence of each of the
        four XRD assumptions (see Sect.~\ref{sec:approx_XRD}) taken
        separately shows that the assumption on absorption profiles
        has a major impact on the $S_{23}$ source
        function. Quantitatively, we measure a mean relative
        difference between the solution thus obtained and the FNLTE
        solution of the order of $65\%$. Qualitatively, and as
        discussed above, this assumption considerably increases the
        $S_{23}$ source function in the areas where the overshoot is
        observed in XRD. However, the amplitude of the
          overshoot is increased by a factor $\approx 2$. This
        hypothesis alone does not fully explain this
        behaviour. Another XRD hypothesis must therefore have the
        opposite effect, by reducing the $S_{23}$ source function near
        the surface and at high frequency. This is precisely the
        qualitative effect of the hypothesis on the coupling between
        the VDFs in the Boltzmann equations. The inclusion of these
        two assumptions alone explains the appearance of this
        overshoot, with a maximum relative deviation between the
        solution thus obtained and the XRD solution of the order of
        $\approx 3\%$. The other assumptions have a much smaller
        impact, of the order of a few percent on average. However, what is more
        important is that the FNLTE approach allows us to get rid of all
        these assumptions and solves the XRD overshooting issue. 
        It is important to note that the XRD overshooting problem seen in the present work (see Fig.~\ref{fig:Fig04_aa55008-25}) is also due to the use of a three-level model atom with infinitely sharp levels. More specifically, the overshoot in $S_{23}$ from the XRD hypothesis is a pathological feature of $R_I$ \citep{hummer62_redistrib} and $P_I$ \citep{Hubeny_2photons} and does not occur with $R_{II}$ and $P_{II}$, which include the natural broadening of energy levels. Indeed, the latter has been verified using the MALI-XRD method of \cite{MALI_XRD}. Moreover, in a more realistic case including more atomic levels, one would expect to observe a strong influence of the higher levels on level 3, modifying the behaviour near the surface.
        
\begin{figure}[t]
                        \includegraphics[width=\columnwidth]{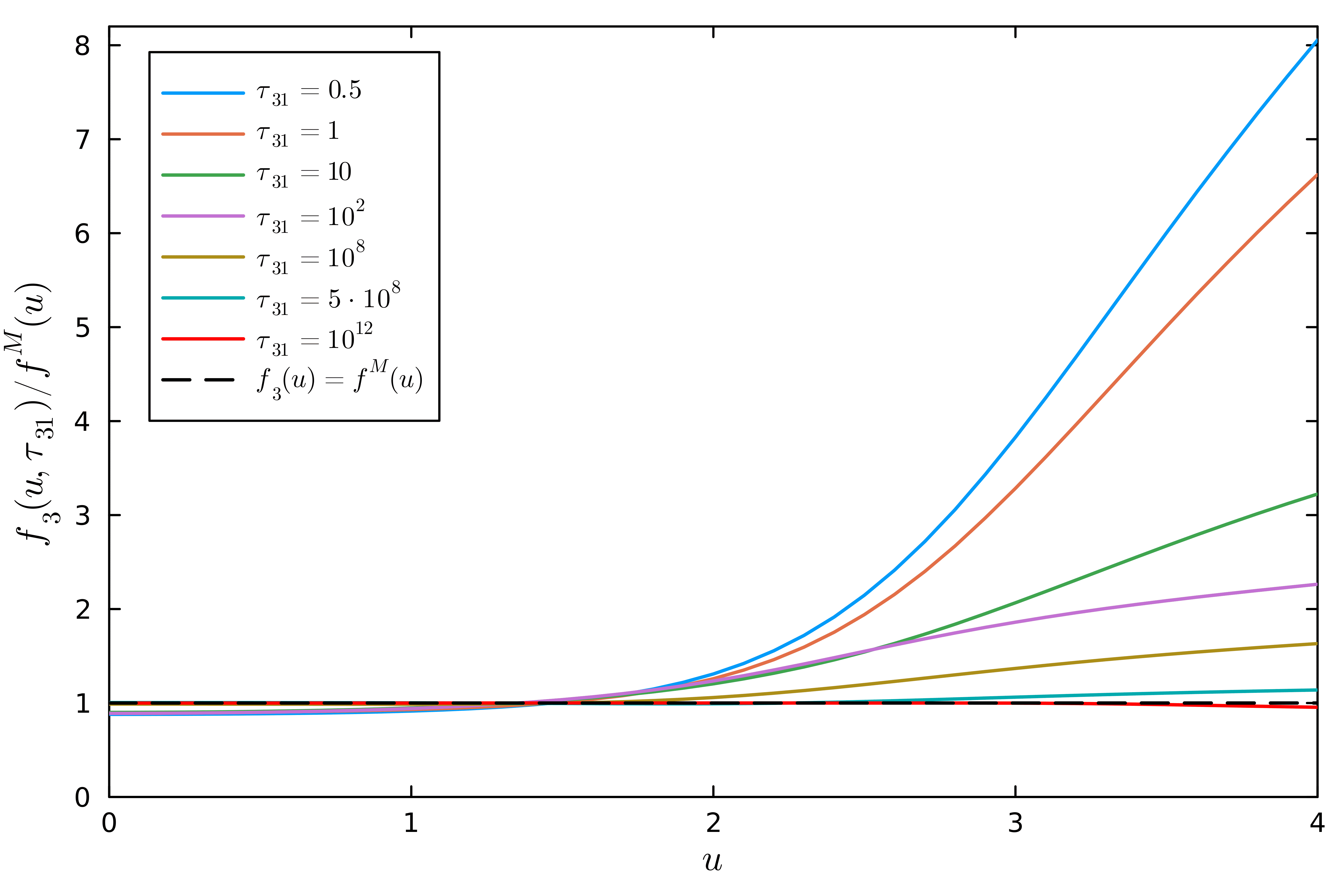}
                        \caption{Normalised atomic velocities variations of
                                 $f_3$ velocity distribution of the second
                                excited level of a three-bound-level hydrogen
                                atom. Results (coloured lines) were obtained for
                                different optical depths: $\tau_{31}=0.5,1,10,10^2,10^8,5\cdot 10^8,10^{12}$. The dashed black line shows the Maxwellian limit $f_3(u)=f^M(u)$.}
                        \label{fig:Fig08_aa55008-25}
                \end{figure}

        Finally, it has been emphasised throughout this paper that the
        FNLTE approach for radiative modelling deals not only with the
        possible deviations of the radiation field and atomic
        populations from their equilibrium distributions, but also for
        the deviation of the VDFs from
        Maxwellian. Therefore, in addition to the usual source
        functions, we are now able to obtain all VDFs, $f_i \neq f^M,$
        of the excited levels. As $S_{21}$, we find that $f_2$ varies
        only slightly from the two-level results previously obtained
        by PSP23. Indeed, we measure a mean relative deviation of
        $0.084\%$ between these two solutions. However, the velocity
        distribution associated with level $3$ is a new feature
        introduced by the FNLTE approach, as displayed in
        Fig.~\ref{fig:Fig08_aa55008-25}. The higher the atomic velocity modulus $u$ is,
        the more $f_3$ deviates from the Maxwellian. In deeper layers,
        as $f_3 \rightarrow f^M$ the atmosphere tends to be
        thermalised. We also note that the variations in $f_3$ are
        very similar to those in $\rho_{23}$ (XRD), which is perfectly
        understandable given the explicit dependence of the profiles
        on the distribution functions (especially because $\psi_{32}$
        is calculated using $f_3$).

        \section{Conclusions}
        \label{sec:conclusion}
                
        Although they were thoroughly formulated more than forty years
        ago by \citealt{HOSI,HOSII} \citep[see also][]{OxeniusBook}, FNLTE effects for a multi-level atom had never been computed
        before the present study. As a first step, we restricted
        ourselves to a simple three-bound-level model atom, based on
        parameters initially provided by \cite{Avrett}, and infinitely
        sharp levels for several benchmark purposes.

        As our preliminary solutions to the multi-level FNLTE problem
        cannot be directly compared with other studies, we verified
        our ability to reproduce several `degraded' problems with
        accuracy. This also led us to point out some inconsistencies of the
        quite advanced still XRD modelling.

        The level of realism of the atomic and scattering models
        will need to be improved in order to take into account
        the natural broadening of excited atomic levels. However,
        the major conceptual jump from two to three or more levels has
        already been made, and the \cite{SPV24} step will be
          generalised further.
        
        In the longer term, we could also consider non-Maxwellian free
        electrons in a self-consistent way as other massive
        particles. Cases where stimulated emission is not negligible
        could also be explored, although this would involve major
        changes in the calculation of atomic profiles.
        
        Finally, we are aware that realism comes at a high computational
        price. For this purpose, we need to develop a more efficient
        numerical scheme. However, another study of ours
        \citep{UBU} shows that it is possible to adapt approximate
        operator methods \citep[see e.g.][]{hubeny2003accelerated,FBF}
        to the two infinitely-sharp-level FNLTE problem. Therefore,
        in the near future, a robust and convergent method similar to these
        ones will be adapted to our new multi-level,
        multi-distribution problem.
                
        \begin{acknowledgements}
                T. Lagache is kindly supported by a `PhD booster'
                grant by the Toulouse Graduate School of Earth and
                Space Sciences (TESS; \url{https://tess.omp.eu/}). Authors thank Prof. Ivan Hubeny for very critical and insightful
                comments.
        \end{acknowledgements}
        
        \bibliographystyle{aa}
        \bibliography{biblio}

\begin{thebibliography}{29}
\expandafter\ifx\csname natexlab\endcsname\relax\def\natexlab#1{#1}\fi

\bibitem[{Avrett(1968)}]{Avrett}
Avrett, E.~H. 1968, in Resonance Lines in Astrophysics, eds. R. G. Athay, J.
  Mathis, A. Skumanich (Boulder: National Center for Atmospheric Research), 27

\bibitem[{Gray(2012)}]{Maser}
Gray, M. 2012, Maser Sources in Astrophysics, Vol.~50 (Cambridge University
  Press)

\bibitem[{Heinzel {et~al.}(1987)Heinzel, Gouttebroze, \&
  Vial}]{heinzel1987_PRD}
Heinzel, P., Gouttebroze, P., \& Vial, J.-C. 1987, A\&A, 183, 351

\bibitem[{Huben{\'y}(1981)}]{Hubeny81}
Huben{\'y}, I. 1981, A\&A, 100, 314

\bibitem[{Huben{\'y}(1982)}]{Hubeny_2photons}
Huben{\'y}, I. 1982, JQSRT, 27, 593

\bibitem[{Huben{\'y}(1985)}]{hubeny1985ETLA}
Huben{\'y}, I. 1985, Bulletin of Astronomical Institutes of Czechoslovakia, 36,
  1

\bibitem[{Huben{\'y}(2003)}]{hubeny2003accelerated}
Huben{\'y}, I. 2003, Stellar Atmosphere Modeling, 288, 17

\bibitem[{Huben{\'y} \& Cooper(1986)}]{Hubeny_Cooper1986}
Huben{\'y}, I. \& Cooper, J. 1986, ApJ, 305, 852

\bibitem[{Huben{\'y} \& Lites(1995)}]{hubeny_lites}
Huben{\'y}, I. \& Lites, B.~W. 1995, ApJ, 455, 376

\bibitem[{Huben{\'y} \& Mihalas(2014)}]{HMbook}
Huben{\'y}, I. \& Mihalas, D. 2014, Theory of stellar atmospheres: An
  introduction to astrophysical non-equilibrium quantitative spectroscopic
  analysis, Vol.~26 (Princeton University Press)

\bibitem[{Huben{\'y} \& Oxenius(1987)}]{Hubeny_3photons}
Huben{\'y}, I. \& Oxenius, J. 1987, JQSRT, 37, 65

\bibitem[{Huben{\'y} {et~al.}(1983{\natexlab{a}})Huben{\'y}, Oxenius, \&
  Simonneau}]{HOSI}
Huben{\'y}, I., Oxenius, J., \& Simonneau, E. 1983{\natexlab{a}}, JQSRT, 29,
  477

\bibitem[{Huben{\'y} {et~al.}(1983{\natexlab{b}})Huben{\'y}, Oxenius, \&
  Simonneau}]{HOSII}
Huben{\'y}, I., Oxenius, J., \& Simonneau, E. 1983{\natexlab{b}}, JQSRT, 29,
  495

\bibitem[{Hummer(1962)}]{hummer62_redistrib}
Hummer, D.~G. 1962, MNRAS, 125, 21

\bibitem[{Hummer(1969)}]{hummer69_solutions}
Hummer, D.~G. 1969, MNRAS, 145, 95

\bibitem[{Lagache {et~al.}(2025)Lagache, Sampoorna, \& Paletou}]{UBU}
Lagache, T., Sampoorna, M., \& Paletou, F. 2025, RAS Techniques and
  Instruments, rzaf023

\bibitem[{Milkey {et~al.}(1975)Milkey, Shine, \& Mihalas}]{milkey1975resonance}
Milkey, R.~W., Shine, R.~A., \& Mihalas, D. 1975, ApJ, 199, 718

\bibitem[{Oxenius(1965)}]{Oxenius65}
Oxenius, J. 1965, JQSRT, 5, 771

\bibitem[{Oxenius(1986)}]{OxeniusBook}
Oxenius, J. 1986, Kinetic theory of particles and photons: theoretical
  foundations of non-LTE plasma spectroscopy (Berlin: Springer)

\bibitem[{Paletou(1995)}]{MALI_PRD}
Paletou, F. 1995, A\&A, 302, 587

\bibitem[{Paletou \& Auer(1995)}]{FBF}
Paletou, F. \& Auer, L.~H. 1995, A\&A, 297, 771

\bibitem[{Paletou \& Lagache(2025)}]{MALICRD_HAL}
Paletou, F. \& Lagache, T. 2025, {\url{https://hal.science/hal-05051883}}

\bibitem[{{Paletou} \& {L{\'e}ger}(2007)}]{Paletou_Leger2007}
{Paletou}, F. \& {L{\'e}ger}, L. 2007, JQSRT, 103, 57

\bibitem[{Paletou \& Peymirat(2021)}]{PP21}
Paletou, F. \& Peymirat, C. 2021, A\&A, 649, A165

\bibitem[{Paletou {et~al.}(2023)Paletou, Sampoorna, \& Peymirat}]{PSP23}
Paletou, F., Sampoorna, M., \& Peymirat, C. 2023, A\&A, 671, A93

\bibitem[{Rybicki \& Hummer(1991)}]{MALI}
Rybicki, G.~B. \& Hummer, D.~G. 1991, A\&A, 245, 171

\bibitem[{Sampoorna {et~al.}(2013)Sampoorna, Nagendra, \& Stenflo}]{MALI_XRD}
Sampoorna, M., Nagendra, K.~N., \& Stenflo, J.~O. 2013, ApJ, 770, 92

\bibitem[{Sampoorna {et~al.}(2024)Sampoorna, Paletou, Bommier, \&
  Lagache}]{SPV24}
Sampoorna, M., Paletou, F., Bommier, V., \& Lagache, T. 2024, A\&A, 690, A213

\bibitem[{Uitenbroek(1989)}]{Uitenbroek_1989}
Uitenbroek, H. 1989, A\&A, 213, 360

\end{thebibliography}
        
        \begin{appendix}
                
                \section{Derivation of observer's frame profiles}
                \label{app:App1}
                
        The absorption and emission profiles in the observer's frame
        are given by the Eqs.~(\ref{eq:phi1})
        and~(\ref{eq:psi1}). For infinitely sharp levels, we
        have given the form of the atomic profiles in the
        Eqs.~(\ref{eq:prof_ato_abs})
        and~(\ref{eq:prof_ato_em}). Thus, with $j > i$, we
        write (for example for $\psi$ but the calculation is
        completely analogous for $\varphi$):
        \begin{equation}
                        \psi_{ji}(\nu_{ji}) = \oint
                        \frac{d\Omega_{ji}}{4\pi} \int f_j(\vec{u})
                        \delta(\xi_{ji} - \nu_{0,ij}) d^3\vec{u} \,.
        \end{equation}
        Using the definitions of the reduced frequency $x_{ji}$, the
        normalised velocity $u$ and the Doppler-Fizeau effect given in
        Eq.~(\ref{eq:Doppler}), we obtain:
                \begin{equation}
                        \psi_{ji} = \frac{1}{\Delta_{ji}} \oint
                        \frac{d\Omega_{ji}}{4\pi} \int f_j(\vec{u})
                        \delta(x_{ji} - \vec{u} \cdot
                        \vec{\Omega_{ji}}) u^2 du d\Omega_u \,.
                \end{equation}
                Noticing that:
                \begin{equation}
                        \oint \frac{d\Omega_{ji}}{4\pi} \delta(x_{ji}
                        - \vec{u} \cdot \vec{\Omega_{ji}}) =
                        \frac{1}{2u} H(u-\lvert x_{ji} \rvert) \,,
                \end{equation}
                with $H$ the Heaviside function, then:
                \begin{equation}
                        \psi_{ji}(x_{ji}) = \frac{1}{2
                          \Delta_{ji}} \int_{\lvert x_{ji}
                          \rvert}^{\infty} f_j(u) u du \,.
                \end{equation}

                \section{Derivation of a general expression for emission profiles in the XRD approximations}
                \label{app:App2}
                
                The aim here is to obtain a general expression for the
                emission profiles within the framework of the
                approximations of the so-called XRD approach of the
                non-LTE radiative transfer. These approximations are
                described in detail in
                Sect.~\ref{sec:approx_XRD}. Before applying this set
                of assumptions, we propose a more general formulation
                based on the kinetic equations given in
                Eq.~(\ref{eq:Boltz}). For this purpose, we
                rewrite these equations as:
                \begin{equation}
                        n_i \Pi_i(\vec{u}) f_i = L_i \,,
                        \label{eqAppB:Boltz}
                \end{equation}
                with,
                \begin{equation}
                        \Pi_i(\vec{u}) = Q_{V,i} + \sum_{j \neq i}
                        \left[ \tilde{R}_{ij}(\vec{u}) + C_{ij}
                          \right] \,,
                \end{equation}
                and, following HOSI, we introduce $L_i$, the density
                corresponding to the ensemble of processes
                populating naturally or not the atomic level $i$ in
                the phase space:
                \begin{equation}
                        L_i = n_i Q_{V,i} f^M + \sum_{j \neq i} n_j
                        f_j \left[ \tilde{R}_{ji}(\vec{u}) + C_{ji}
                          \right] \,.
                \end{equation}
                
                Analogously, $\Pi_i(\vec{u})$ is related to the
                density corresponding to the set of processes
                depopulating the level $i$ in phase
                space\footnote{Velocity-changing collisions cannot
                populate a given level $i$ from a level $j$. However,
                they change the velocity of the atom inducing a
                transition from one state (position and velocity) to
                another in phase space. These collisions therefore
                populate a level by creating or destroying particles in
                phase space.}. In the previous equations,
                $\tilde{R}_{ij}(\vec{u})$ is the radiative rate
                dependent on the velocity associated with the
                transition $ i \rightarrow j$. It is defined as:
                \begin{equation}
                        \tilde{R}_{ij}(\vec{u}) = \left\{ 
                        \begin{aligned}
                                B_{ij} J_{ij}(\vec{u}) \qquad
                                \text{if}\,\,i<j\,,\\ A_{ij} \qquad
                                \text{if}\,\,i>j\,.
                        \end{aligned}
                        \right.
                \end{equation}
                Integration over all velocities of this quantity,
                i.e.:
                \begin{equation}
                        R_{ab} = \int d^3\vec{u} f_a(\vec{u})
                        \tilde{R}_{ab}(\vec{u}) \,,
                \end{equation}
                leads to the standard radiative rate which appear in
                the IKEE defined as:
                \begin{equation}
                        R_{ij} = \left\{
                        \begin{aligned}
                                B_{ij}\mathcal{J}_{ij} \qquad
                                \text{if}\,\,i<j\,,\\ A_{ij} \qquad
                                \text{if}\,\,i>j\,.
                        \end{aligned}
                        \right.
                \end{equation}
                Also, with velocity-changing collisions, we define $P_i$ as:
                \begin{equation}
                        P_i = Q_{V,i} + \sum_{j \neq i} (R_{ij} + C_{ij}) \,,
                \end{equation}
                where stimulated emission is neglected. Then, we can
                write the $\beta_{ji}$ atomic emission profiles
                describing the $j \rightarrow i$ transition (this
                formula was adapted from HOSI; see
                also our Sect.~\ref{sec:prof_ato}) as:
                \begin{equation}
                        \begin{split}
                                \beta_{ji} &=
                                \mathrm{prob}(\rightarrow j^*) r_{ji}
                                + \sum_{l <j}
                                \mathrm{prob}(\rightarrow l^*
                                \Rightarrow j)j_{lji} \\ & +
                                \sum_{k<l<j} \mathrm{prob}(\rightarrow
                                k^* \Rightarrow l \Rightarrow j)
                                j_{klji} \\ & + \sum_{m<k<l<j}
                                \mathrm{prob}(\rightarrow m^*
                                \Rightarrow k \Rightarrow l
                                \Rightarrow j)j_{mklji} + ...
                        \end{split}
                \end{equation}
                Next, according to HOSI, we can write:
                \begin{equation}
                        \mathrm{prob}(\rightarrow j^*) =
                        \frac{1}{L_j}\left[ n_jf^MQ_{V,j} + \sum_{p
                            \neq j} n_pf_p (A_{pj} + C_{pj})\right]
                        \,.
                \end{equation}
                The term in the numerator relates to all processes
                that naturally populate level $j$. From the
                denominator, we retrieve the term $L_j$, the meaning
                of which we have already discussed. Also, the
                probability that level $j$ is non naturally populated
                from a level $l<j$ is\footnote{For $l=1,\,2$,
                we get Eqs (7.4), (7.7) and
                (7.8) of HOSI.}:
                \begin{equation}
                        \mathrm{prob}(\rightarrow l^* \Rightarrow j) =
                        \mathrm{prob}(\rightarrow l^*) \times \frac{
                          n_lf_lB_{lj}J_{lj}(\vec{u}) }{L_j} \,.
                \end{equation}
                We also have:
                \begin{equation}
                        \mathrm{prob}(\rightarrow k^* \Rightarrow l
                        \Rightarrow j) = \mathrm{prob}(\rightarrow k^*
                        \Rightarrow l) \times \frac{
                          n_lf_lB_{lj}J_{lj}(\vec{u}) }{L_j} \,,
                \end{equation}
                and,
                \begin{equation}
                        \begin{split}
                                \mathrm{prob}(\rightarrow m^*
                                \Rightarrow k \Rightarrow l
                                \Rightarrow j) &=
                                \mathrm{prob}(\rightarrow m^*
                                \Rightarrow k \Rightarrow l)
                                \\ & \times \frac{ n_lf_lB_{lj}J_{lj}(\vec{u})
                                }{L_j} \,.
                        \end{split}
                \end{equation}
                
                In Sect.~\ref{sec:FNLTE3}, we have neglected
                all processes involving three or more
                photons. In other words, we have implicitly made the
                assumption that $\mathrm{prob}(\rightarrow k^*
                \Rightarrow l \Rightarrow j)=0$, $\forall k,l,j$ with
                $k<l<j$. We must therefore have $n_lf_lB_{lj}J_{lj}(\vec{u})=0$
                which is absurd as none of these quantities
                can be zero or, $\mathrm{prob}(\rightarrow
                  k^*\Rightarrow l)=0$, implying\footnote{Because
                  $\mathrm{prob}(\rightarrow k^* \Rightarrow l) +
                  \mathrm{prob}(\rightarrow l^*)=1$}
                $\mathrm{prob}(\rightarrow
                l^*)=1$. Thus, to
                satisfy this last condition, we must assume that all
                levels $l$ such that $l<j$ are naturally
                populated. This assumption only applies to the
                calculation of a given emission profile. For example,
                at three levels, the calculation of $\beta_{31}$ and
                $\beta_{32}$ is made by assuming that level $2$ is
                naturally populated (which is factually false because
                radiative absorption $1 \rightarrow 2$ does not lead to
                natural population of level $2$). On the other hand,
                for the calculation of $\beta_{21}$, level $2$
                will not be considered to be naturally populated in
                any circumstances. With this clarification, we write
                the atomic emission profiles as:
                \begin{equation}
                        \beta_{ji} = \mathrm{prob}(\rightarrow j^*)
                        r_{ji} + \sum_{l <j} \mathrm{prob}(\rightarrow
                        l^* \Rightarrow j)j_{lji} \,.
                \end{equation}
                Finally, the emission profile in the observer's frame
                $\psi_{ji}$ now writes, using Eq.~(\ref{eq:psi1}) and
                neglecting velocity-changing collisions, as:
                \begin{equation}
                        \begin{split}
                                \psi_{ji} &= \oint
                                \frac{d\Omega_{ji}}{4\pi} \int
                                d^3\vec{u} \frac{f_j(\vec{u})}{L_j}
                                \left\{ r_{ji} \sum_{p \neq j}
                                n_pf_p(A_{pj} + C_{pj}) \right.\\ &+
                                \left. \sum_{l<j}
                                \mathrm{prob}(\rightarrow l^*) \times
                                n_lf_lB_{lj}J_{lj}(\vec{u}) j_{lji} \right\}
                                \,.
                        \end{split}
                \end{equation}
                To continue, we use the Boltzmann equations
                i.e. Eq.~(\ref{eqAppB:Boltz}) and our assumption on
                the natural population of atomic levels $l<j$, namely,
                $\mathrm{prob}(\rightarrow l^*)=1$; which is a
                consequence of the assumption on processes with three
                or more photons. We then have:
                \begin{equation}
                        \begin{split}
                                \psi_{ji} &= \oint
                                \frac{d\Omega_{ji}}{4\pi} \int
                                \frac{d^3\vec{u}}{n_j \Pi_j(\vec{u})}
                                \left\{ r_{ji} \sum_{p \neq j}
                                n_pf_p(A_{pj} + C_{pj}) \right.\\ &+
                                \left. \sum_{l<j} n_lf_lB_{lj}J_{lj}(\vec{u})
                                j_{lji} \right\} \,.
                        \end{split}
                \end{equation}
                In order to retrieve usual non-LTE transfer
                approximations (including XRD), we use $\Pi_i \approx
                P_i$ (equivalent to assumption (iii) -- see Sect.~\ref{sec:approx_XRD}; see also HOSI for more details), and the above
                equation reduces to:
                \begin{equation}
                        \begin{split}
                                \psi_{ji} &= \frac{1}{n_jP_j} \oint
                                \frac{d\Omega_{ji}}{4\pi} \int
                                d^3\vec{u} \left\{ r_{ji} \sum_{p \neq
                                  j} n_pf_p(A_{pj} + C_{pj})
                                \right.\\ &+ \left. \sum_{l<j}
                                n_lf_lB_{lj}J_{lj}(\vec{u}) j_{lji} \right\}
                                \,.
                        \end{split}
                \end{equation}
                We recall that $j_{lji} = \int I_{lj} r_{lji}
                d\xi_{lj} / J_{lj}(\vec{u})$ and the generalised scattering
                integrals denoted $\mathcal{R}_{lji}$ are defined as:
                \begin{equation}
                        \mathcal{R}_{lji}(x_{ij}) = \oint
                        \frac{d\Omega_{lj}}{4\pi} \int I_{lj}
                        \frac{R^l_{lji}(x_{ij},
                          x_{lj})}{\varphi_{ij}^{M*}} dx_{lj} \,,
                \end{equation}
                with $\varphi_{ij}^{k*}$ the absorption profile in the
                observer's reference frame calculated with the VDFs
                $f_k$ and $R_{lji}^{l}$ the generalised redistribution
                function integrated in angle and velocity (and
                calculated with $f_l$). Following HOSII, they are
                defined respectively as:
                \begin{equation}
                        \varphi_{ij}^{k*} = \oint
                        \frac{d\Omega_{ji}}{4\pi} \int d^3\vec{u}
                        f_k(\vec{u}) r_{ji} \,,
                        \label{eqAppB:phik*}
                \end{equation}
                and,
                \begin{equation}
                        R_{lji}^{l}(x_{ij},x_{lj}) = \oint
                        \frac{d\Omega_{ji}}{4\pi} \int d^3\vec{u}
                        f_l(\vec{u}) r_{lji} \,.
                        \label{eqAppB:rllji}
                \end{equation}
                The emission profile $\psi_{ji}$ is then finally
                written as:
                \begin{equation}
                        \psi_{ji} = \frac{\varphi_{ij}^{M*}}{n_jP_j}
                        \left\{ \sum_{l<j} n_lB_{lj}
                        \mathcal{R}_{lji}^{l} + \sum_{p \neq j}
                        n_p(A_{pj} + C_{pj})
                        \frac{\varphi_{ij}^{p*}}{\varphi_{ij}^{M*}}
                        \right\} \,.
                \end{equation}
                By distinguishing Raman and resonance scattering, and
                noting $\mathcal{R}_{lji}^{l} = \bar{P}_{lji}^l$ and
                $\mathcal{R}_{iji}^{i} = \bar{J}_{iji}^{i}$
                respectively their generalised scattering integrals,
                we have:
                \begin{equation}
                        \psi_{ji} = \frac{\varphi_{ij}^{M*}}{n_jP_j}
                        \left\{ n_iB_{ij}\bar{J}_{iji}^{i} +
                        \sum_{\substack{l\neq i,j \\ l<j }} n_lB_{lj}
                        \bar{P}_{lji}^l \right.\\ \left. + \sum_{p
                          \neq j} n_p(A_{pj} + C_{pj})
                        \frac{\varphi_{ij}^{p*}}{\varphi_{ij}^{M*}}
                        \right\} \,.
                        \label{eq:more_general_psi}
                \end{equation}
                Finally, we make the last assumption of the XRD
                approximation: to calculate $f_i$, we assume that
                $f_j=f^M$ $\forall j\neq i$. Thus, the quantities
                defined in Eqs.~(\ref{eqAppB:phik*})
                and~(\ref{eqAppB:rllji}) are calculated using a
                Maxwellian VDF and we simply obtain, after
                manipulating sums and indices:
                \begin{equation}
                        \begin{split}
                                \psi_{ji} = &
                                \frac{\varphi_{ij}^{M*}}{n_j P_j}
                                \left\{ n_i B_{ij} \bar{J}_{iji}^M +
                                \sum_{\substack{l\neq i,j \\ l<j }}
                                n_l B_{lj} \bar{P}_{lji}^M
                                \right. \\ & \left. + \left[ n_iC_{ij}
                                  +\sum_{k \neq i,j} n_k (A_{kj} +
                                  C_{kj}) \right]
                                \vphantom{\sum_{\substack{l\neq i,j
                                      \\ l<j }}} \right\} \,,
                        \end{split}
                \end{equation}
                which is completely equivalent to
                Eq. (\ref{eq:prof_em_general_formula_samp}).
                
                \section{Benchmarking against standard PRD}
                \label{app:App3}

\begin{figure}
                        \includegraphics[width=\columnwidth]{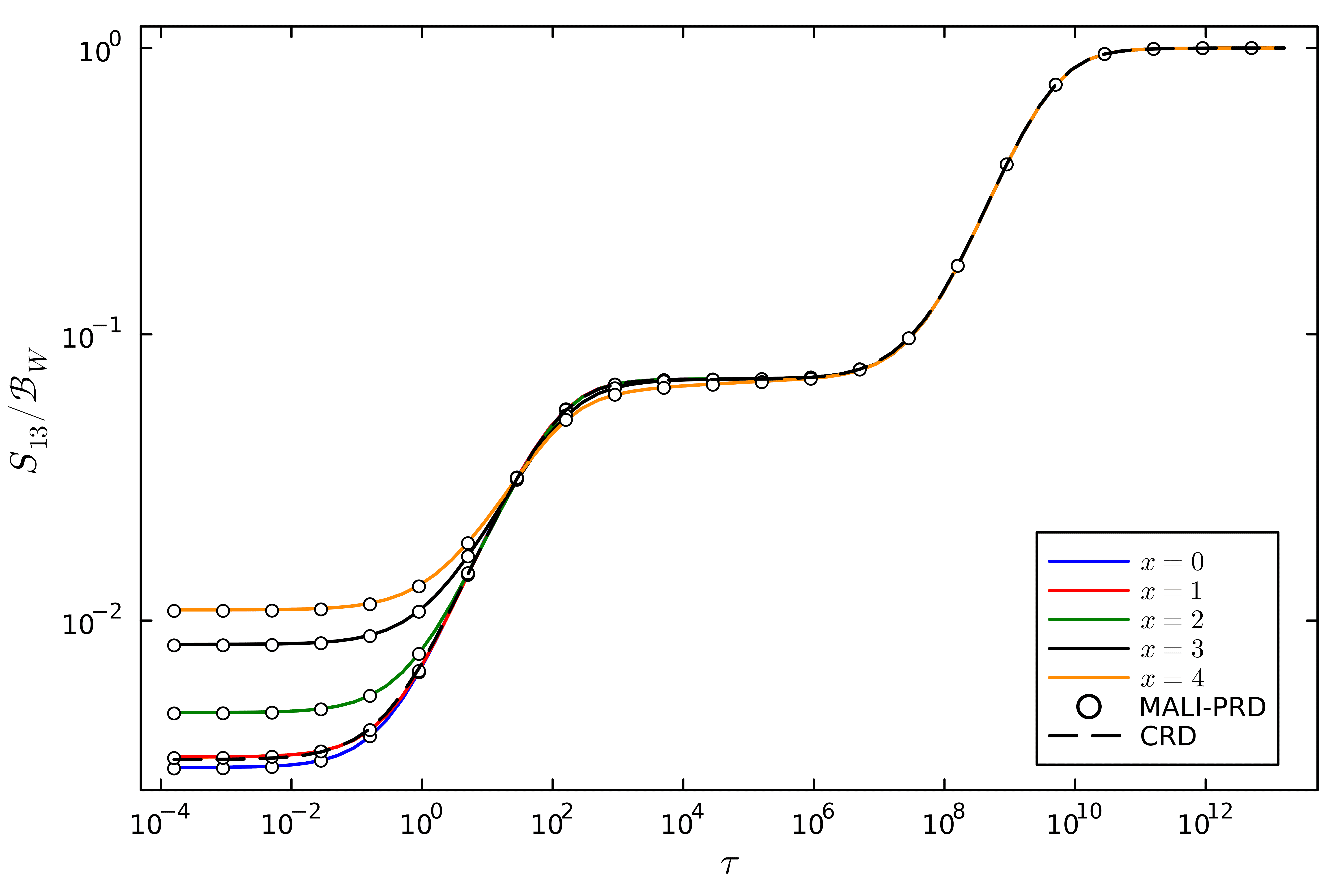}
                        \caption{Optical depth variations of
                          normalised source functions for 
                          $1 \leftrightarrow 3$ line of a three-level hydrogen
                          atom. We compare FNLTE results (coloured
                          lines) for various frequencies $x=0,1,2,3,4$
                          with PRD reference solutions (black open
                          circle) computed using the MALI-PRD method
                          of \cite{MALI_PRD}. The black dashed line
                          shows the CRD solution computed using
                          MALI-CRD method.}
                        \label{fig:FigC01_aa55008-25}
                \end{figure}

\begin{figure}
                        \includegraphics[width=\columnwidth]{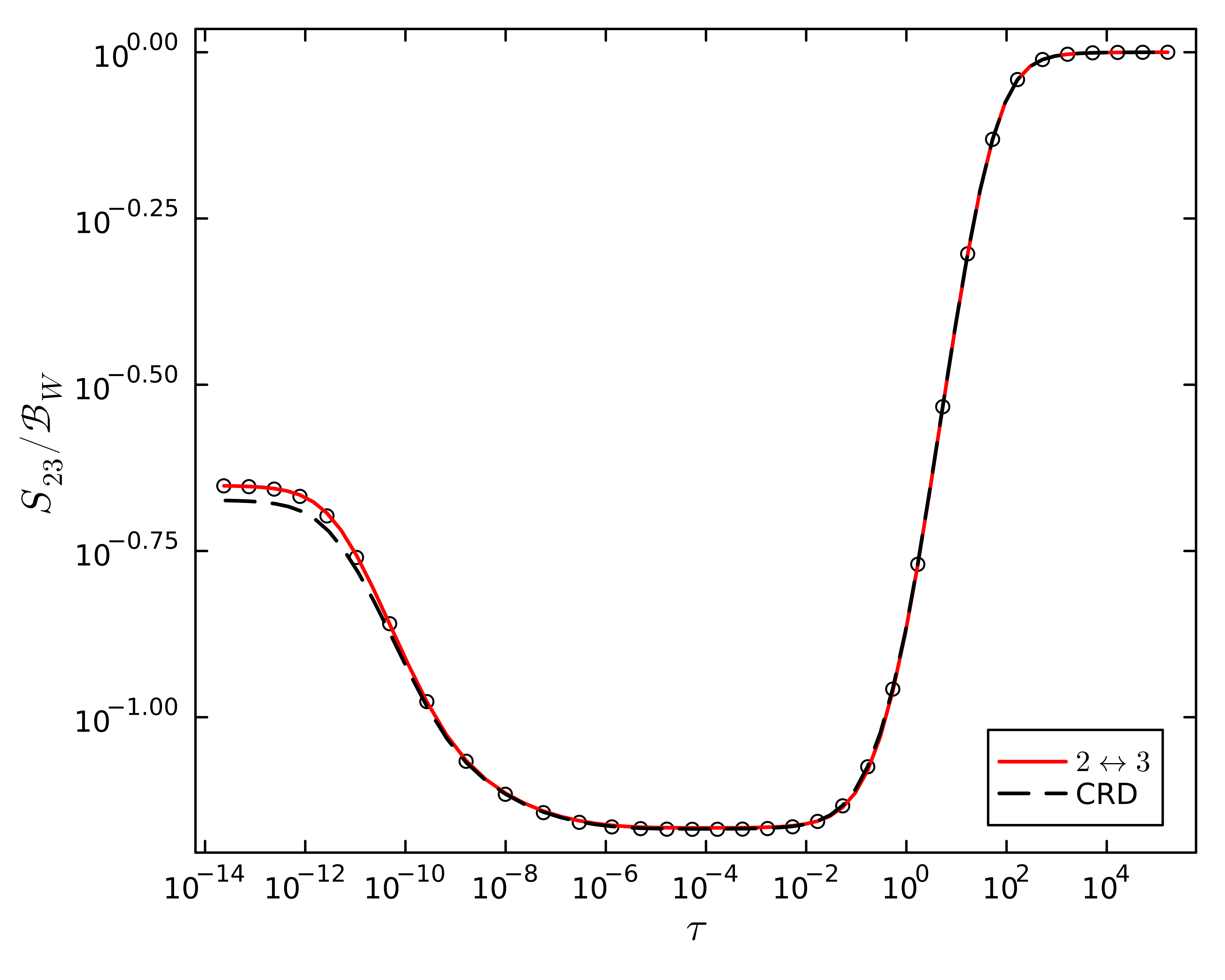}
                        \caption{Optical depth variations of
                          normalised source functions for
                          $2 \leftrightarrow 3$ line of a three-level hydrogen
                          atom. We compare FNLTE results (red line)
                          with PRD reference solutions (black open
                          circle) computed using the MALI-PRD method
                          of \cite{MALI_PRD}. The black dashed line
                          shows the CRD solution computed using
                          MALI-CRD method.}
                        \label{fig:FigC02_aa55008-25}
                \end{figure}

                We are fully aware that cross-redistribution
                (XRD) is a mechanism
                still rarely invoked in radiative modelling. As an
                extra validation test, we propose here to study the
                case of standard Hummer's PRD, which is absolutely
                central to explaining the observed properties of the
                $1 \leftrightarrow 2$ and $1 \leftrightarrow 3$ lines of the Sun
                \citep[see][]{hubeny_lites,heinzel1987_PRD}. In the
                present case, we consider a hydrogen atom with three infinitely sharp levels
                 for which only Hummer's PRD effects will
                be considered for the two resonance lines ($1 \leftrightarrow 2$ and $1 \leftrightarrow 3$ lines), and where the CRD
                approximation will be used for $2 \leftrightarrow 3$ line
                i.e. $\psi_{32} = \varphi_{23}$. For $1 \leftrightarrow 2$ and
                $1 \leftrightarrow 3$ lines, an emission profile is calculated without
                taking into account the effects of
                cross-redistribution. This is equivalent to
                considering the second term of
                Eq.~(\ref{eq:prof_em_general_formula_samp}) as zero,
                i.e. $\bar{P}^M_{lji}=0$. For $1 \leftrightarrow 2$ line, this
                consideration has no impact and so $f_2$ will always
                be given by Eq.~(\ref{eq:Boltz_2_XRD}). On the other
                hand, for $1 \leftrightarrow 3$ line, we must now calculate $f_3$ with
                $J_{23}(\vec{u})=\mathcal{J}_{23}=0$. Note, moreover, that
                according to the IKEE, we have:
                \begin{equation}
                        n_1C_{13} + n_2C_{23} = n_3P_3 -
                        n_2B_{23}\mathcal{J}_{23} -
                        n_1B_{13}\mathcal{J}_{13} \,,
                \end{equation}
                and we finally write:
                \begin{equation}
                        f_3 = f^M + \frac{f^M}{n_3P_3} \Bigg[
                          n_1B_{13}(J_{13}(\vec{u}) - \mathcal{J}_{13}) \Bigg] \, .
                        \label{eq:Boltz_3_PRD}
                \end{equation}
                The combination of
                Eqs.~(\ref{eq:Boltz_2_XRD}),~(\ref{eq:Boltz_3_PRD})
                then allows us to compare the results given by our
                FNLTE multi-level code with the reference results
                calculated using the MALI-PRD method developed by
                \cite{MALI_PRD}. The results are shown in
                Fig.~\ref{fig:FigC01_aa55008-25} for the $1 \leftrightarrow 3$ line and in
                Fig.~\ref{fig:FigC02_aa55008-25} for $2 \leftrightarrow 3$ line. We do not
                show the results for $1 \leftrightarrow 2$ line for the same
                reasons explained in
                Sect.~\ref{sec:approx_XRD}. Finally, we see that the
                expected results are very well reproduced, with a
                mean relative error between the two solutions
                equal, for $1 \leftrightarrow 2$, $1 \leftrightarrow 3$ and $2 \leftrightarrow 3$ lines
                respectively, to about $0.22 \%$, $0.37 \%$ and $0.34
                \%$.
                
        \end{appendix}
        
\end{document}